\documentclass[aps,prb]{revtex4-2}
\usepackage{bm}
\usepackage{amsmath}
\usepackage{color}
\usepackage{xcolor}
\usepackage{amssymb}
\usepackage[shortlabels]{enumitem}
\usepackage{graphicx}
\usepackage[colorlinks=true,linkcolor=blue,citecolor=blue]{hyperref}
\usepackage{subcaption}
\usepackage{caption}
\usepackage{physics}
\DeclareMathAlphabet{\mathbbmsl}{U}{bbm}{m}{sl}
\makeatletter
\newsavebox{\@brx}
\newcommand{\llangle}[1][]{\savebox{\@brx}{\(\m@th{#1\langle}\)}%
	\mathopen{\copy\@brx\kern-0.5\wd\@brx\usebox{\@brx}}}
\newcommand{\rrangle}[1][]{\savebox{\@brx}{\(\m@th{#1\rangle}\)}%
	\mathclose{\copy\@brx\kern-0.5\wd\@brx\usebox{\@brx}}}
\makeatother

\begin{document}

\draft

\title{Surface effects on the plasmons in two-dimemsional heterostructures: Application to tilted semi-Dirac materials}

\author{Teresa Lee$^{1,2}$, Godfrey Gumbs$^{1,2}$, Thi Nga Do$^{3}$,  Andrii Iurov$^{2,4}$ and Danhong Huang$^{5}$  }
 \address{$^1$Department of Physics, Hunter College, City University of New York, 695 Park Avenue, New York, NY 10065, USA}
\address{$^{2}$The Graduate School and University Center, The City University of New York, New York, NY 10016, USA}
\address{$^{3}$Department of Physics and Astronomy, The University of Tampa, Tampa, Florida 33606, USA}
\address{$^{4}$Department of Physics and Computer Science, Medgar Evers College of City University of New York, Brooklyn, NY 11225, USA} 
\address{$^{5}$Space Vehicles Directorate, US Air Force Research Laboratory, Kirtland 
Air Force Base, New Mexico 87117, USA}

\date{\today}

\begin{abstract} 
We derive closed-form analytic expressions for the plasmon dispersion equations  for two and three monolayers embedded in a non-uniform dielectric medium with a surface. There is no electron tunneling between the layers or hybridization of the layer polarizability. The dispersion equations are deduced from the calculated expressions for the corresponding surface response functions (SRFs) generated by a frequency-dependent external polarized electromagnetic field. The SRF is calculated by employing Maxwell’s equations in conjunction with linear response theory. The dispersion functions reduce to well-known results for two and three monolayers embedded in a bulk medium with a uniform dielectric background. We examine the role played by a surface (i.e., homogeneous versus inhomogeneous dielectric background screening), for gapped tilted semi-Dirac materials (TSDMs) with half-linear, half-parabolic spectrums whose energy bands are tilted, anisotropic in wave vector space, and a gap is induced. The number of plasmon branches is always equal to the number of monolayers. However, the separation between these branches depends on the distance between the layers and crucially on the chosen dielectric material between the layers and whether there is air or a substrate surrounding the layered structure. For gapped TSDM, the Landau damping, i.e., Plasmon lifetime,  varies from branch to branch as well as along the branch for a chosen wave vector direction. We examine these novel behaviors for two and three gapped TSDM monolayers assuming different values for the dielectric background.  Our results could be useful for comparing theory with experimental data from electron energy loss spectroscopy (EELS) data.
\end{abstract}

\date{\today}

\maketitle

\medskip

\medskip

\noindent
{\bf  Corresponding author}:\ \     Teresa Lee;   E-mail:   
Teresa.Lee80@login.cuny.edu

\section{Introduction}
\label{sec1}

Since the reported novel experimental results on the electron transport properties of monolayer graphene which generated an enormous amount of theory and experiment on this material, several unique exotic Dirac and Weyl structures  have been investigated \cite{Intro1}. These include silicene \cite{Wu2016,Intro2,Intro2-2}, phosphorene \cite{Intro3} and germanene \cite{Intro4}.  For these, their electronic, optoelectronic, electron transport  and  optical properties have been the subject of numerous investigations. In most Dirac materials, the relativistic massless quasiparticles described by the Dirac Hamiltonian have an isotropic linear spectrum forming a symmetric Dirac cone in momentum space. However, it has recently been shown that some systems instead display anisotropic linear spectra characterized by tilted Dirac cones. Examples include, 8-Pmmn borophene and the 1T's phase of monolayer transition-metal dichalcogenides (TMDCs) \cite{Sadhukhan, Balassis}. Now, 8-Pmmn borophene is, a stable 2D monolayer, semi metallic allotrope of boron, characterized by eight boron atoms in its unit cell and a distinctive Pmmn space group. Renowned for possessing unique, highly anisotropic and tilted Dirac cones, it is a promising material for next-generation, high mobility electronics valleytronics and specialized optoelectronic applications \cite{Lopez}. 
\medskip
\par

Concerning the anisotropic materials, numerous studies have focused on semi-Dirac materials, where the intrinsic low-energy bands of these monolayer structures exhibit zero band gap with a half-linear, half-parabolic spectrum \cite{Andrii,Yan}. Conveying its exotic property of massless in one direction and massive in other, semi-Dirac materials started to be extensively studied in both experiment and theoretical research studies. One experimental result suggests the existence of its dispersion in potassium-doped black phosphorus, with energy dispersion linear along the armchair direction and parabolic along the zigzag direction at a specific dopant density \cite{jkim}. Moreover, various experimental techniques were used to investigate its dispersion, such as fabrication of planar microcavities etched on honeycomb lattices of micropillars which led to the observation of  anisotropic transport in polaritons \cite{Real}. On the other hand, semi-Dirac dispersion has been investigated in many theoretical research works including strained honeycomb lattice \cite {Dietl}, and multi-layered nanostructures of $VO_{2}$ layers confined within $TiO_2$ \cite{Banerjee,Pardo}. These researches include analysis of distinctive anisotropic properties such as anisotropic transport of polaritons,  plasmon excitations\cite{Andrii, Real}, and further studies of semi Dirac materials in various contexts are ongoing \cite{Delplace, Kush, Kotov,Elsayed}.
\medskip
\par

In view of the comments we made above concerning  the current interest of both experimentalists and theoreticians in anisotropic semi-Dirac materials, we will use gapped and tilted semi Dirac materials(TSDMs) for demonstrating with two and three layers the important differences in the plasmon dispersions made by a homogeneous (bulk related) versus inhomogeneous (surface related) dielectric background.  We would like to make it clear that there are already theoretical studies on the plasmonic properties of single or double layers of TSDMs \cite{Andrii, Debasmita}. However, in those works, the monolayers are embedded in a uniform bulk background dielectric material. To go beyond these works, we calculate the surface response function (SRF) of a layered structure where the 2D layers could be on the surface as a protective coating exposed to  air. We chose gapped TSDM since the energy  bands have multi degrees of freedom such as tilting  and anisotropy, but the formalism is not restricted to these materials. In this paper, we derive the plasmon dispersion function for multi-layered two-dimensional (2D) structures. The layers are separated by dielectric materials which are not necessarily homogeneous in the case of three layers suspended in vacuum. The plasmon dispersion for three layers then can be used to deal with two layers with one end exposed to air and the other end placed on a dielectric substrate.

\medskip
\par

Here, we obtained an exact solution of the plasmon dispersion equation in the random-phase-approximation (RPA) with the anisotropic density-density response function of a multi-layered 2D system \cite{JainAllen,Andrii2}. Our method of calculation is to use the SRF for layers of electrons with chosen dielectrics around them \cite{Dipendra}. There is no electron tunneling between layers or hybridization of the polarization function. We analytically derive formulas for the dispersion relations in the long wavelength limit to display the plasmon dispersions, and also use numerical calculations to display density plots of the loss functions for arbitrary wavelength. These results reveal the anisotropy of the modes, and the density plots for two and three layers particularly show the acoustic and optical modes having different dampings.

\medskip
\par

The remainder of this paper is organized as follows. In Sec.~\ref{sec2}, we describe, review and analyze the Hamiltonian for TSDMs with and without the gaps arising from an underlying dielectric material, such as hexagonal boron nitride, or from spin-orbit coupling (SOC). We calculate the eigenvalues and eigenfunctions, and investigate the energy dispersions of TSDMs in two perpendicular momentum space directions with chosen parameters. In Sec.~\ref{sec3}, we derive and analyze the SRF for multiple layers, starting from three layers enclosing two different dielectric media between them and freely suspended in vacuum. We then deduce the SRF for a two-layered structure enclosing a dielectric medium and exposed to air on one end and an arbitrary dielectric substrate on the other end of the structure. We quantitatively compare the plasmon dispersions for a two-layered structure with this inhomogeneous dielectric background with the case when these two layers are embedded in a homogeneous material. Lastly, we consider a monolayer of gapped TSDM placed on a substrate. In Sec.~\ref{sec4}, we calculate the polarization function which is employed in the SRF to determine the plasma excitation dispersions for multiple layers of gapped TSDMs in the long wavelength limit, as well as in general. We qualitatively and quantitatively compare the plasmon excitation energies, the separation between branches and anti-crossing of these branches. We also compare their dampings as well as how this damping varies along a branch, and the way in which these behaviors are affected by the chosen dielectric background. In Sec.~\ref{sec5}, we calculate the absorption coefficients. Sec.~\ref{sec6} summarizes the main results of this paper and discusses potential applications to sensing, computing and energy through the manipulation of light on the nanoscale. This can be achieved by combining light waves with collective charge density oscillations, i.e., plasmons.

\section{Model Hamiltonian for Tilted semi-Dirac material}
\label{sec2}

\subsection{Tilted gapless semi-Dirac Hamiltonian}

We begin by introducing and reviewing the low-energy model Hamiltonian for a monolayer thick TSDM which will be exploited below in our studies of the plasmon excitations in a layered structure embedded either in a homogeneous or inhomogeneous dielectric background. This Hamiltonian is quadratic in the momentum $k_{x}$ direction and linear in the perpendicular $k_{y}$ direction and is given by \cite{Yan}

\begin{equation} 
\hat{H}_{\xi}(\textbf{{k}}|a,\tau) = \xi\tau\hbar\upsilon_{F}k_{y}\hat{\Sigma}_{0}^{(2)}+\hbar a k_{x}^{2}\hat{\Sigma}_{x}^{(2)}+\hbar\upsilon_{F}k_{y}\hat{\Sigma}_{y}^{(2)} \  ,
\label{eq2}
 \end{equation}
where $ \hat{\Sigma}_{0}^{(2)},  \hat{\Sigma}_{x}^{(2)}, 
 \hat{\Sigma}_{y}^{(2)}$ are Pauli matrices. In the first term of Eq.~(\ref{eq2}), $\tau$ is a tilting parameter for the energy bands, and is defined as the ratio between the Fermi velocity along the  tilting direction and the Fermi velocity without tilting, i.e., $\upsilon_{t}/\upsilon_{F}$. Also, $\xi$ labels the valley. We note that there is $\upsilon_{F}$ multiplying every wave vector component $k_{y}$. This $\hbar\upsilon_{F}k_{y}$ form indicates that graphene-like linear property exists in the $k_{y}$ direction, and that we are applying tilting in the $k_{y}$ direction only. In contrast, there exists a quantity $a=\hbar/2m^\ast$ in the second term. This $\frac{\hbar^{2}k_{x}^{2}}{2m^\ast}$ form implies that, there exists a nonlinear form for the energy in the $k_x$-direction. In matrix form, the Hamiltonian is expressed as

\begin {equation}
\hat{H}_{\xi}(k|a,\tau) =\left(
\begin{array}{cc} 
\xi\hbar\tau\upsilon_{F}k_{y} & \hbar a k_{x}^{2}-i\hbar\upsilon_{F}k_{y} \\
\hbar ak_{x}^{2}+i\hbar\upsilon_{F}k_{y} &\xi\hbar\tau\upsilon_{F}k_{y} \end{array}
\right)
\end{equation}
which readily yields the anisotropic eigenvalues for tilted gapless semi-Dirac materials, i.e.,

\begin{equation}
\begin{split}
\varepsilon_{\lambda,\xi}(\textbf{k}|a,\tau)
=\hbar\xi\tau\upsilon_{F}k_{y}+\lambda\sqrt{(\hbar\upsilon_{F}k_{y})^2+(\hbar ak_{x}^2)^2}
\end{split}\  .
\label{GG1}
\end{equation}
The corresponding wave function is

\begin{equation}
\ket{\mathbf{k},\lambda|a\mathbf{\tau}}=\frac{1}{\sqrt{2}}
\left(
\begin{array}{cc}
1\\
\lambda\frac{\sqrt{(\hbar\upsilon_{F}k_{y})^{2}+(\hbar ak_{x}^{2})^{2}}}{\hbar ak_{x}^{2}-i\hbar\upsilon_{F}k_{y}}
\end{array}\right)
\end{equation}
which can be expressed as shown below in terms of the  angle $\Theta(\mathbf{k}|a)=\tan^{-1}(\frac{\upsilon_{F}k_{y}}{ ak_{x}^{2}})$

\begin{equation}
\ket{\mathbf{k},\lambda|a\mathbf{\tau}}=\frac{1}{\sqrt{2}}
\left[
\begin{array}{cc}
1\\
\lambda e^{i\Theta(\mathbf{k}|a)}
\end{array}\right]
\end{equation}

\begin{figure}[h]
\centering
\begin{subfigure}{.4\textwidth}
\refstepcounter{subfigure}
\centering
\includegraphics[width=.9\columnwidth]{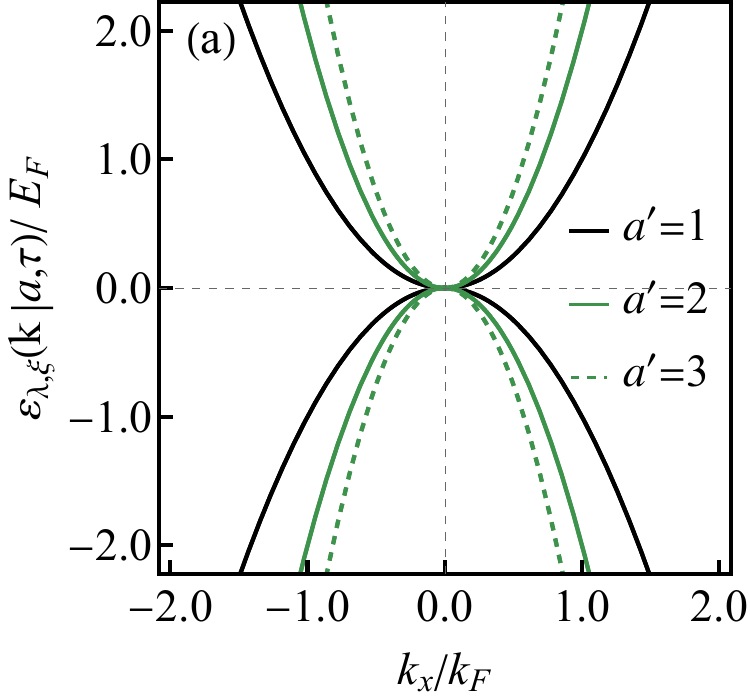}
\caption*{}
\label{sf1} 
\end{subfigure}%
\begin{subfigure}{.4\textwidth}
\refstepcounter{subfigure}
\centering
\includegraphics[width=.89\columnwidth]{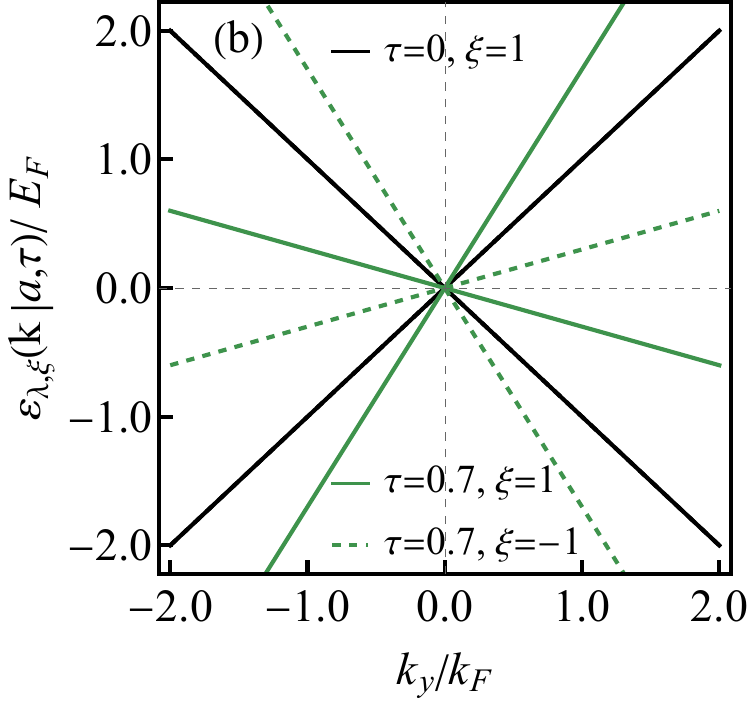}
\caption*{}
\label{sf2}
\end{subfigure}
\captionsetup{singlelinecheck=off,justification=raggedright}\caption{(Color online) (a) Energy dispersion $\varepsilon_{\lambda,\xi}(\textbf{k}|a,\tau)/E_{F}$  as a function of wave vector $k_{x}/k_{F}$ when the component $k_{y} $=0 and the tilting $\tau$=$0$. Here, $E_{F}$=$\hbar \upsilon_{F}k_{F}$ is the Fermi energy of graphene. $a^\prime$=$\frac{a}{a_{F}}$, where $a_{F}$ is defined as $\frac{\upsilon _{F}}{k_{F}}$. In (b), the energy dispersion $\varepsilon_{\lambda,\xi}(\textbf{k}|a,\tau)/E_{F}$  as a function of $k_{y}/k_{F}$ when $k_{x}$=0. Each curve is plotted for chosen $a^\prime$ and $\tau$.}
\label{figure1}
\end{figure}
\medskip
\par
Figure~\ref{figure1} shows the energy dispersion derived from the tilted semi-Dirac Hamiltonian when either $k_{x}$ or $k_{y}$ is set equal to zero. The energy is in unit of $E_F$ for graphene $\hbar\upsilon_{F}k_{F}$. Figure~\ref{sf1} presents the energy dispersion plotted as a function of $k_{x}/k_{F}$ when $k_{y} =0$. We define an inverse effective mass parameter $a^\prime$ as $a/a_{F}$, the inverse effective mass in unit of $a_{F}$, which came from the relation $\hbar k_{F}^{2}a_F=\hbar\upsilon_{F}k_F$, thus $a_{F}=\upsilon_{F}/k_F$ of graphene, and we plot the energy dispersion for three different $a^\prime$. As presented in Fig.~\ref{sf1}, when $k_{y}=0$, the energy dispersion becomes a quadratic function of $k_{x}$. There is no tilting via  $\tau$. However, when $a^\prime$ is increased from $1.0\times10^{-3}$ to $3.0\times 10^{-3}$, the energies become narrower. In contrast, when $k_{x}=0$, the energy dispersion is linear in $k_{y}$ as shown  in Fig.~\ref{sf2}. This dispersion depends on the tilting parameter $\tau$, and as $\tau$ is increased from 0 to 0.7, we observe that the energy dispersion with $\xi$ = 1.0 is tilted counterclockwise, while the energy dispersion with $\xi$ =-1.0 is tilted in the clockwise direction. We note that in both cases, there is no energy gap between the valence and conduction bands.

\subsection{Gapped Tilted Semi-Dirac Energy Bands}

As we mentioned above, since our goal is to investigate thoroughly how the plasmon excitation energies, their dampings as well as the separation between the branches  in a multi-layer structure are all determined  by the surrounding dielectric background, we must consider the case when the TSDM energy bands have a gap. Such a gap can be accomplished through a substrate or SOC. For this, we add a term $\Delta\hat{\Sigma}_{z}^{(2)}$ to the Hamiltonian in Eq.~(\ref{eq2}), where $\Delta$ could represent the magnitude of the SOC or half of the band gap, and $\hat{\Sigma}_{z}^{(2)}$ is a Pauli matrix, i.e.,

\begin {equation}
\hat{H}_{\xi}(\textbf{{k}}|a,\tau) = \xi\tau\hbar\upsilon_{F}k_{y}\hat{\Sigma}_{0}^{(2)}+\hbar ak_{x}^{2}\hat{\Sigma}_{x}^{(2)}+\hbar\upsilon_{F}k_{y}\hat{\Sigma}_{y}^{(2)}+\Delta\hat{\Sigma}_{z}^{(2)}\  .
 \end{equation}
 The eigenvalues for the gapped TSDM are now given by

\begin{eqnarray} 
\varepsilon_{\lambda,\xi}(\textbf{k}|a,\tau,\Delta) =& \xi\tau\hbar\upsilon_{F}k_{y}+\lambda\sqrt{(\hbar\upsilon_{F}k_{y})^{2}+(\hbar ak_{x}^{2})^{2}+\Delta^{2}}
\nonumber\\
\approx&
\xi\tau\hbar\upsilon_{F}k_{y}+\lambda\Delta  \left\{1+\frac{(\hbar\upsilon_F)^2}{2\Delta^2} k_y^2+
\frac{(\hbar a)^2}{2\Delta^2} k_x^4-
\frac{(\hbar \upsilon_F)^4}{8\Delta^4} k_y^4
\cdots
\right\}\  .
\label{approxGG}
 \end{eqnarray}
We emphasize that this expansion explicitly assumes that $\Delta$ is finite. The corresponding eigenfunction is 

\begin{equation}
\ket{\mathbf{k},\nu}=\frac{1}{\sqrt{2+\frac{\Delta^{2}}{(\hbar\upsilon_{F}k_{y})^{2}+(\hbar ak_{x}^{2})^{2}}}}
\left(
\begin{array}{cc}
1\\
\lambda\frac{\sqrt{(\hbar\upsilon_{F}k_{y})^{2}+(\hbar ak_{x}^{2})^{2}+\Delta^{2}}}{\hbar ak_{x}^{2}-i\hbar\upsilon_{F}k_{y}}
\end{array}\right)
\end{equation}

\begin{figure}[h]
\centering
\begin{subfigure}{.4\textwidth}
\refstepcounter{subfigure}
\centering
\includegraphics[width=.9\columnwidth]{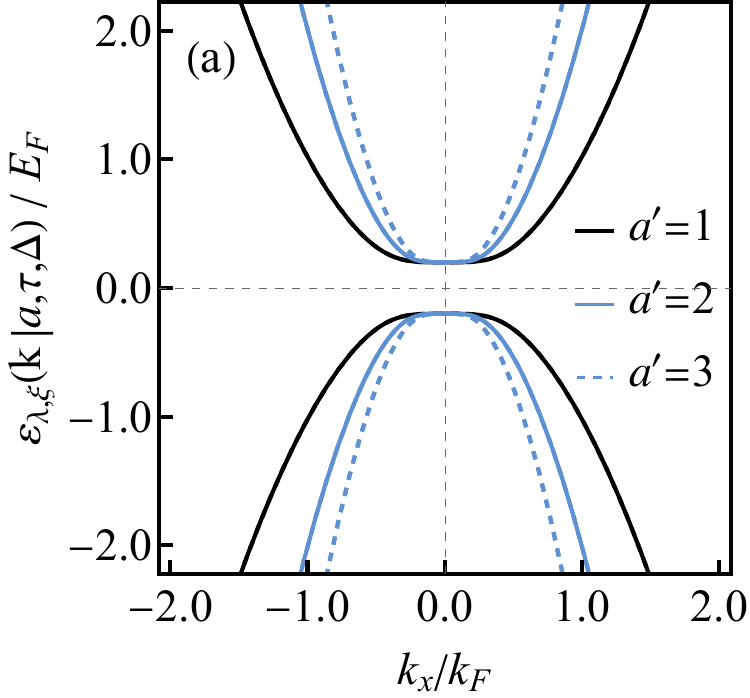}
\caption*{}
\label{sf3} 
\end{subfigure}%
\begin{subfigure}{.4\textwidth}
\refstepcounter{subfigure}
\centering
\includegraphics[width=.89\columnwidth]{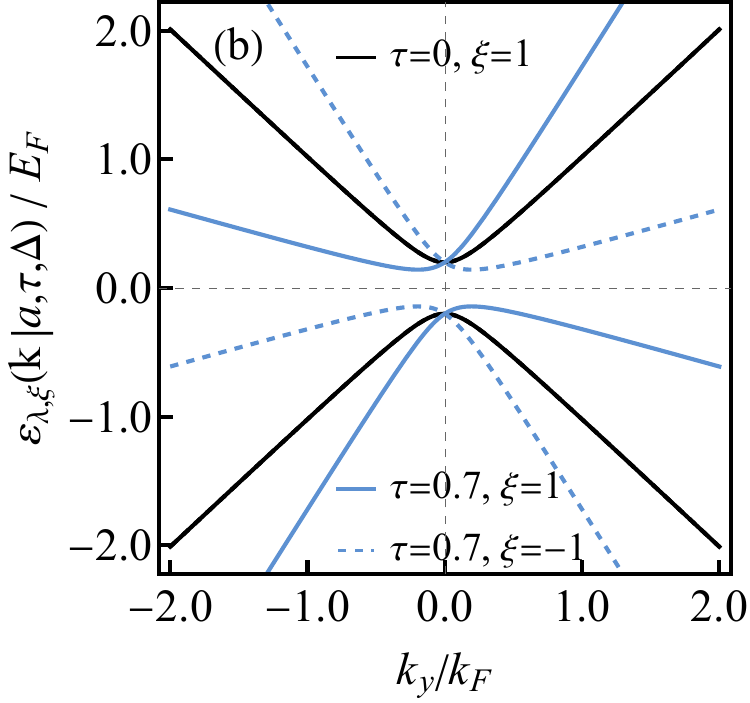}
\caption*{}
\label{sf4}
\end{subfigure} \captionsetup{singlelinecheck=off,justification=raggedright}
\caption{(Color online) (a) Energy eigenvalue $\varepsilon_{\lambda,\xi}(\textbf{k}|a,\tau,\Delta)/E_{F}$ as a function of wave vector $k_{x}/k_{F}$ when $k_{y} $=0. The band gap parameter is $\Delta'$=0.2, where $\Delta^\prime$=$\Delta/E_{F}$ and ${E_{F}}=\hbar\upsilon_{F}k_{F}$, the Fermi energy of graphene. It is plotted for three chosen inverse effective mass parameters $a'$. (b) The energy dispersion $\varepsilon_{\lambda,\xi}(\textbf{k}|a,\tau,\Delta)/E_{F}$ plotted with respect to $k_{y}/k_{F}$ when $k_{x} $=0 and $\Delta^\prime$=0.2. It is presented for two chosen tilting parameters $\tau=0$, and $\tau=0.7$, as well as for two different valley parameters $\xi=1$ and $\xi=-1$.}\  .
\label{figure2}
\end{figure}

\medskip
\par

Figure~\ref{figure2} shows the energy dispersion when a gap is present. The magnitude of energy gap parameter $\Delta^\prime$ which is defined as $\Delta$/$E_{F}$ is set equal to $0.2$. As shown in Fig.~\ref{sf3}, when the energy dispersion is plotted with respect to $k_{x}$/$k_{F}$, we see the same magnitude of energy gaps appearing for different magnitudes of inverse effective mass parameter $a^\prime$. Here, since tilting only takes effect in the $k_{y}$ direction, it does not affect the energy dispersion in the $k_{x}$ direction. In Fig.~\ref{sf4}, as the energy dispersions are plotted as a function of $k_{y}$/$k_{F}$, the gaps appear, and as $\tau$ is increased from 0 to 0.7, the energy dispersion with $\xi$ = 1 is tilted counterclockwise, while the energy dispersion with $\xi$ = -1.0 is tilted in the clockwise direction.

\medskip
\par

\begin{figure}[h]
\centering
\begin{subfigure}{.25\textwidth}
\refstepcounter{subfigure}
\includegraphics[width=0.9\columnwidth]{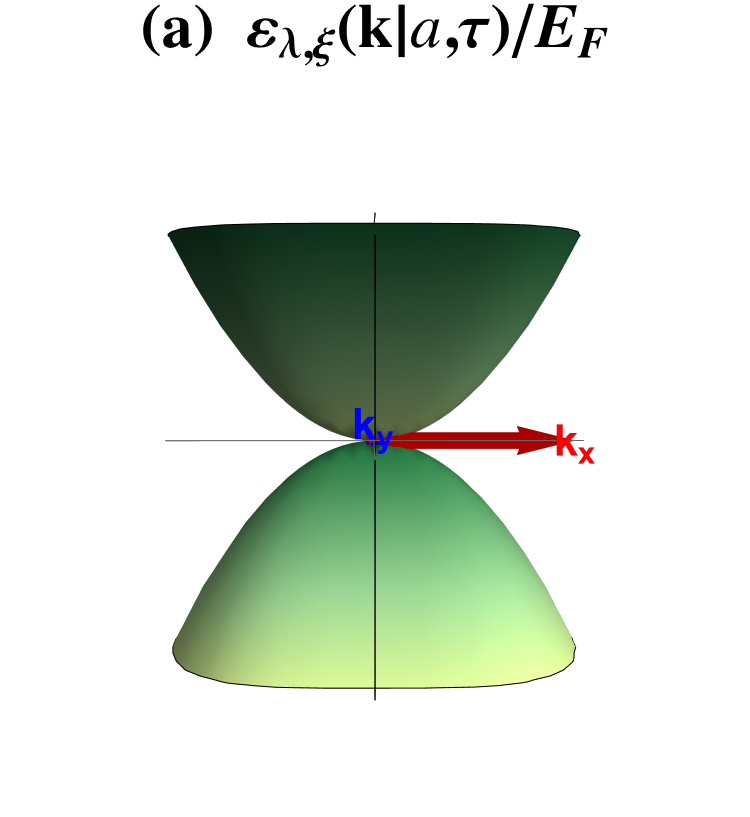}
\caption*{}
\label{E3_4}
\end{subfigure}
\hspace{0.03\textwidth}
\begin{subfigure}{.25\textwidth}
\refstepcounter{subfigure}
\includegraphics[width=0.9\columnwidth]{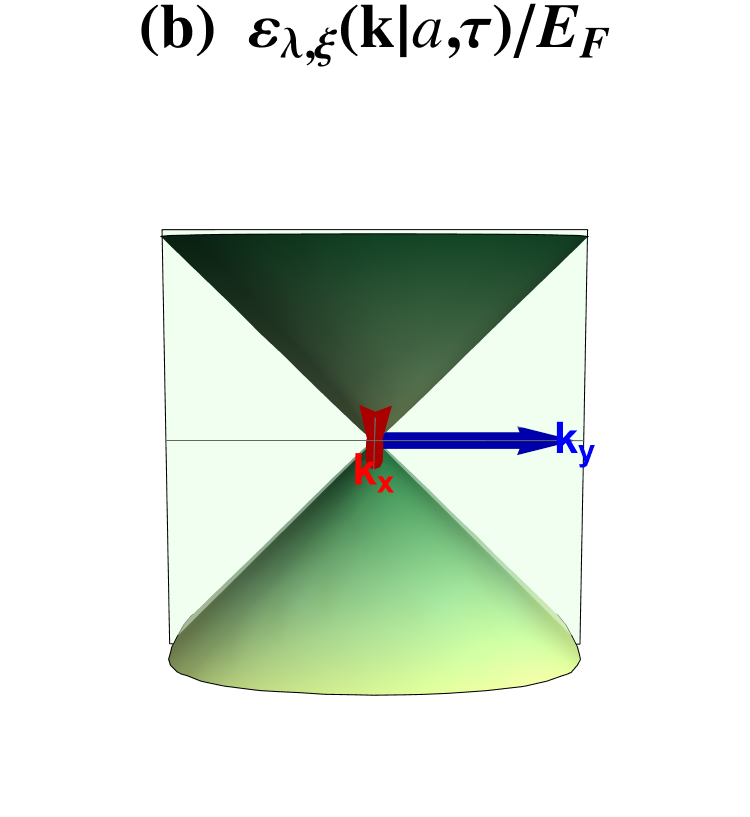}
\caption*{}
\label{E3_5}
\end{subfigure}
\hspace{0.03\textwidth}
\begin{subfigure}{.25\textwidth}
\refstepcounter{subfigure}
\includegraphics[width=0.9\columnwidth]{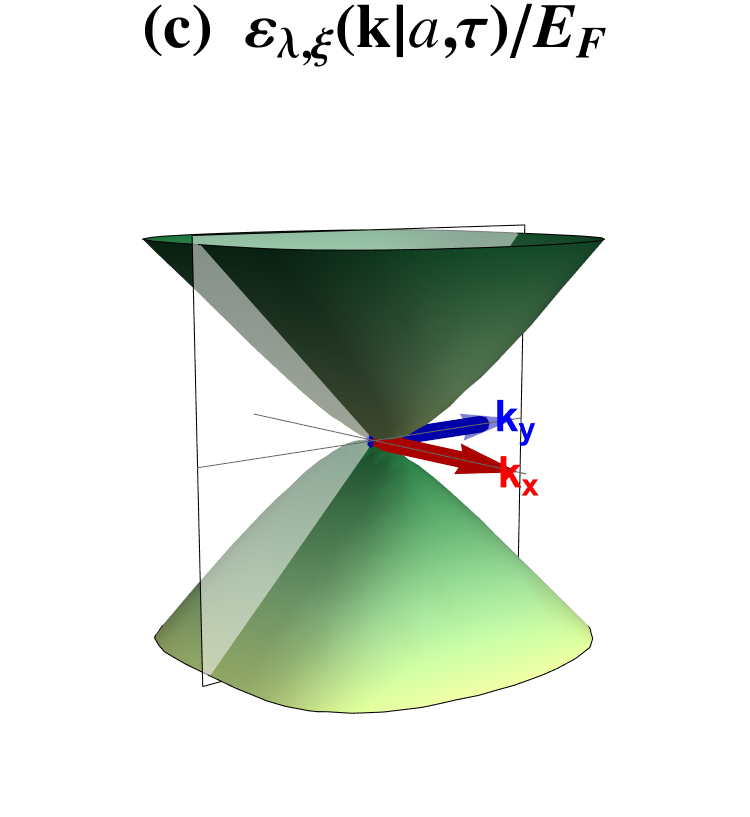}
\caption*{}
\label{E3_6}
\end{subfigure}
\begin{subfigure}{.25\textwidth}
\refstepcounter{subfigure}
\includegraphics[width=0.9\columnwidth]{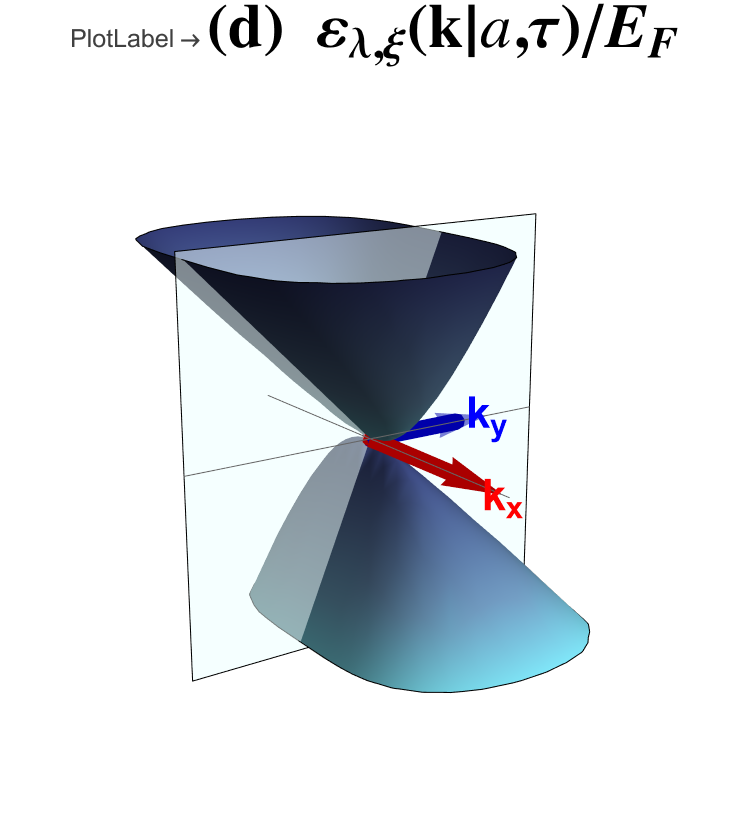}
\caption*{}
\label{E3_4}
\end{subfigure}
\hspace{0.03\textwidth}
\begin{subfigure}{.25\textwidth}
\refstepcounter{subfigure}
\includegraphics[width=0.9\columnwidth]{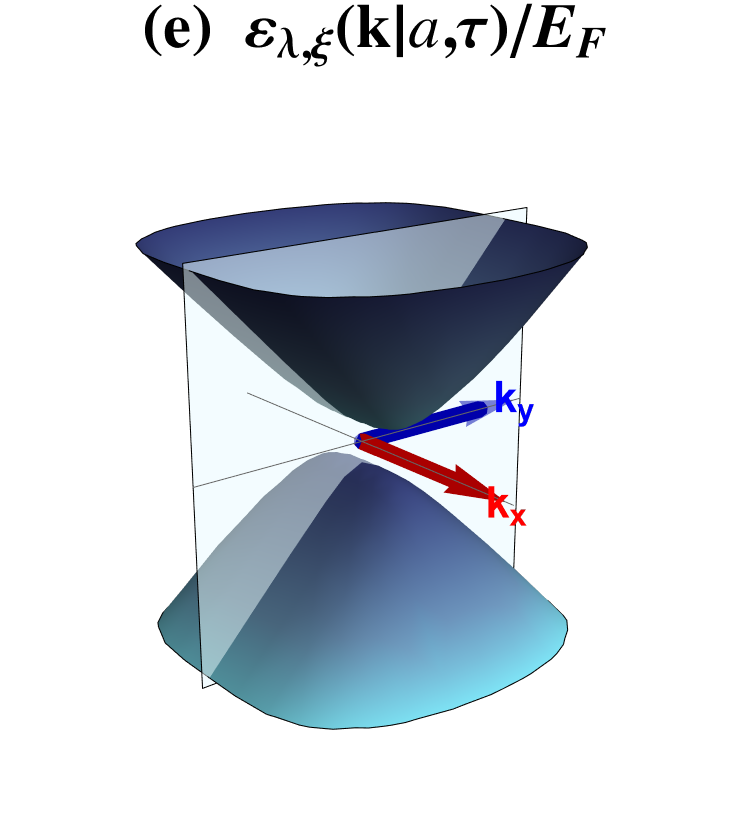}
\caption*{}
\label{E3_5}
\end{subfigure}
\hspace{0.03\textwidth}
\begin{subfigure}{.25\textwidth}
\refstepcounter{subfigure}
\includegraphics[width=0.9\columnwidth]{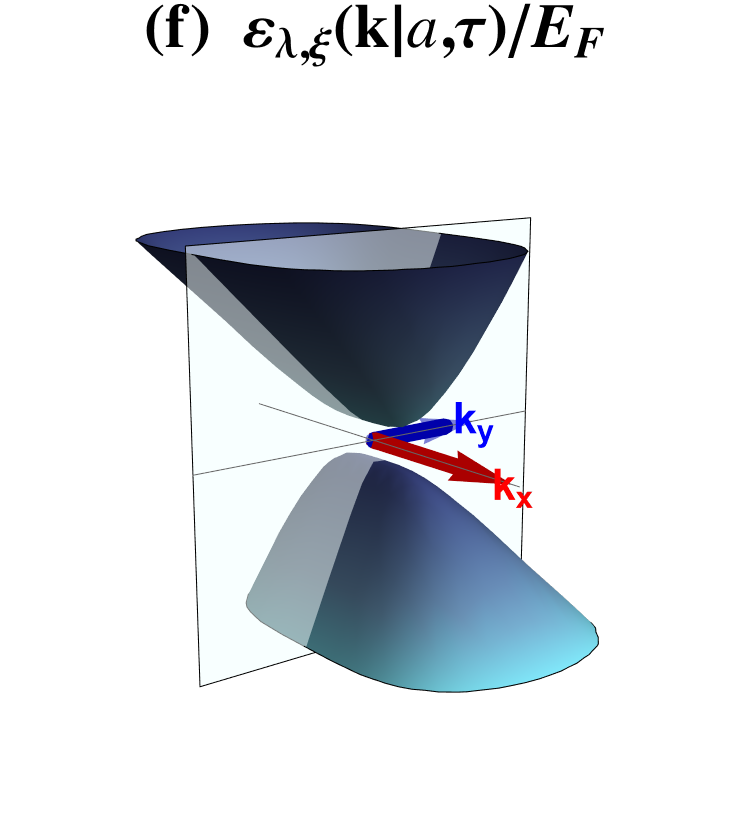}
\caption*{}
\label{E3_6}
\end{subfigure}
\captionsetup{singlelinecheck=off,justification=raggedright}
\caption{(Color online) Three-dimensional view of the energy dispersion in unit of $E_{F}=\hbar\upsilon_{F} k_{F}$. The semi-transparent  $k_{y}$-$\varepsilon_{\lambda, \xi}({\bf k}|a,\tau)$ plane is a geometrical reference plane and not part of the energy dispersion. Energy dispersions with tilting $\tau=0.0$ and energy gap parameter $\Delta^{\prime}=0.0$ are shown in plots (a), (b) and (c). In plot (a), the line of view is along the direction with parabolic dispersion. In plot (b), the line of view is along the direction with linear dispersion. The view in (c) is at an arbitrary angle. Plot (d) shows the energy dispersion where tilting $\tau=0.4$ is present. Plot (e) depicts the energy dispersion when an energy gap parameter $\Delta^{\prime}=0.1$ is present. Plot (f) shows the energy dispersion when both tilting $\tau=0.4$ and a gap parameter $\Delta^{\prime}=0.1$ are present.}
\label{energy3d}
\end{figure}

\newpage
\section{surface response function formalism}
 \label{sec3}
 
In our formalism for calculating the SRF of multi-layered media, we employ Maxwell’s equation $\nabla \cdot \left( {\bf E}({\bf r}) \epsilon_b({\bf r}) \right) =\frac{\rho({\bf  r})}{\epsilon_0}$ where $\epsilon_b({\bf r})$ is the contribution to the dielectric background due to bound electrons which does not depend on the dopant density. Additionally, $\rho(\bf {r})$ denotes the induced electron density fluctuation by an external perturbation such as an impinging beam of electrons from an electron energy loss (EELS) experiment. The resonances in the SRF correspond to plasmon poles and for this reason, we turn our attention  to the following calculations.

\subsection{Three monolayers separated by two dielectric media and freely suspended in vacuum}

\begin{figure}[h]
\captionsetup{singlelinecheck=off,justification=raggedright}
\centering
\includegraphics[width=.5\columnwidth]{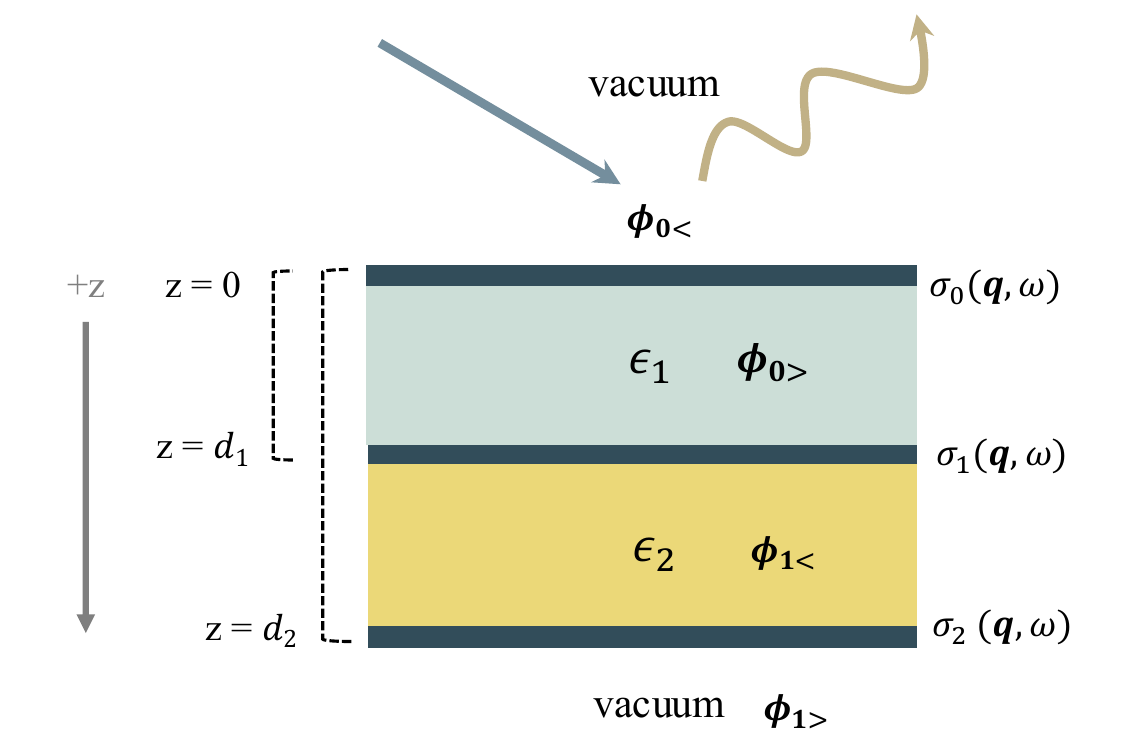}
\caption{(Color online) Schematic representation of three freely suspended electron monolayers with arbitrary separation between pairs. The induced surface charge densities $\sigma_{0}({\bf q},\omega)$, $\sigma_{1}({\bf q},\omega)$ and $\sigma_{2}({\bf q},\omega)$ at z= $0$, z=$d_{1}$ and $z=d_{2}$, respectively, are produced by an incoming beam of electrons scattering off the surface at $z=0 $. The layers are separated by two materials with dielectric constant $\epsilon_{1}$ and $\epsilon_{2}$.}
\label{three-layer}
\end{figure}

First, let us consider an external perturbation incident on the surface lying in the x-y plane of condensed matter. This beam arises from an electromagnetic field, the source of which could be due to an impinging beam of electrons, for example. We now explicitly define and derive the dynamical SRF of frequency $\omega$ and in-plane wave vector ${\bf q}$ due to this perturbation for a heterostructure consisting of three 2D layers. These layers are located at $z=0, d_1$ as well as $z=d_2$ and are translationally invariant in the x-y plane. There is vacuum in the regions $z<0$ and $z>d_2$ but dielectric media fill the spaces {\em between\/} the 2D layers. As Fig.~\ref{three-layer} shows, dielectric materials with constants $\epsilon_1$ and $\epsilon_2$ fill the regions $0\leq z\leq d_1$ and $d_1\leq z\leq d_2$, respectively. Therefore, the 2D layers on the two surfaces are surrounded by vacuum which is unlike the case when the 2D layers are embedded in a material with a uniform background dielectric constant. Under these conditions, the electrostatic potential within the vacuum regions ($z<0$ and $z>d_2$) and the dielectric media $(0<z<d_2)$ must satisfy Gauss’ law for the electrostatic potential $\Phi$, i.e., $\nabla\cdot\left(-\epsilon_b(z)\nabla\Phi({\bf r}_\parallel, z; \omega)\right)=\frac{\rho({\bf r}_\parallel, z;\omega)}{\epsilon_{0}}$, where $\epsilon_b(z)$ represents the spatial distribution  of the background dielectric media and $\rho({\bf r}_\parallel, z;\omega)$ is the induced electron density  on a 2D layer. (The subscript $\parallel$ denotes the in-plane component of a vector and will be dropped when it is convenient to do so.) Furthermore, this may be rewritten, using $\Phi({\bf r})=\phi(z)e^{i{\bf q}_\parallel\cdot{\bf r}_\parallel}$, as $\frac{\partial}{\partial z} \left( \epsilon_b(z)\frac{\partial}{\partial z} \phi(z; {\bf q},\omega)\right) -q^2\phi(z; {\bf q},\omega) =-\frac{\sigma(z; {\bf q},\omega)}{\epsilon_{0}}$ where we took advantage of the in-plane translational symmetry and introduced the planar induced density $\sigma(z; {\bf q},\omega)$. In particular, in the vacuum region {\em close\/} to $z=0$ just above this plane and in the region between $z=0$ and $z=d_1$, the induced electrostatic potential can be expressed as

\begin{equation}
\begin{split}
&\phi_{0<}(z)=e^{-qz}-g({\bf q},\omega)e^{qz},\  \  \  \ z\lesssim 0
\\
&\phi_{0>}(z)=a_{0}e^{-qz}+b_{0}e^{qz}, \  \  \  \  0<z< d_1 \\
\end{split}
\end{equation}
where the coefficients $a_0,b_0 $ and the SRF $g({\bf q},\omega)$ are determined from the continuity equations  $\phi_{0<}(z=0)=\phi_{0>}(z=0) $ as well as  the discontinuity conditions 

\begin{equation}
\begin{split}
&\epsilon_1\  \left.\frac{d\phi_{0>}(z)}{dz}\right|_{z=0} 
-	\left.\frac{d\phi_{0<}(z)}{dz}\right|_{z=0} =-\frac{\sigma_0(z; {\bf q},\omega)}{\epsilon_{0}}  , 
\end{split}
\label{TL1}
\end{equation}
where we take the induced density within the monolayer at $z=0$ to be $\sigma_0(z; {\bf q},\omega)=\delta(z)e^2\chi^{(0)}({\bf q},\omega)\phi_{0>}(z=0)$ which is expressed in terms of the density-density response function  $\chi^{(0)}({\bf  q},\omega)$. After a straightforward calculation, we obtain the following expressions for the coefficients $a_0$, and $b_0$

\begin{equation}
\left(
\begin{array}{cc}
a_{0}\\
b_{0}
\end{array}
\right)
=
 \frac{1}{2\epsilon_1  }
\left(
\begin{array}{cc}
 \epsilon_1+\frac{e^2}{\epsilon_{0}q}\chi^{(0)}({\bf  q},\omega)   & 1 \\\\
 \epsilon_1-\frac{e^2}{\epsilon_{0}q}\chi^{(0)}({\bf  q},\omega)   & -1 
\end{array}
\right)
\left(
\begin{array}{cc}
1-g({\bf q},\omega)\\
1+g({\bf  q},\omega)
\end{array}
\right)
\label{START}
\end{equation}
which  explicitly gives the result for the ratio

\begin{equation}
\frac{b_0}{a_0}=\frac{(1-g({\bf q},\omega))(\epsilon_{1}-\frac{e^2}{\epsilon_0q}\chi^{(0)}({\bf q},\omega))-(1+g({\bf q},\omega))}{(1-g({\bf q},\omega))(\epsilon_1+\frac{e^2}{\epsilon_0q}\chi^{(0)}({\bf q},\omega))+(1+g({\bf q},\omega))}\ .
\label{fracb0a0}
\end{equation}
which will be employed below.
Also, writing in the regions between the second and third monolayers where $d_1<z< d_2$ and $z\gtrsim d_2$,  respectively, for the electrostatic  potentials 

\begin{equation}
\begin{split}
&\phi_{1<}(z)=a_{1}e^{-qz}+b_{1}e^{qz},\  \  \  d_1<z\  <d_2\\
&\phi_{1>}(z)=t_{1}e^{-qz}\  ,  \  \  z\gtrsim  d_2\  ,\\
\end{split}
\label{match2}
\end{equation}  
we match these potentials at $z=d_1$ and $z=d_2$ and take into account the discontinuity of their derivatives across each plane:

\begin{equation}
\begin{split}
\nonumber
&\epsilon_2\  \left.\frac{d\phi_{1<}(z)}{dz}\right|_{z=d_{1}} 
-	\epsilon_1\left.\frac{d\phi_{0>}(z)}{dz}\right|_{z=d_{1}} =-\frac{\sigma_1(z; {\bf q},\omega)}{\epsilon_{0}}  , \\
\nonumber \\
&\left.\frac{d\phi_{1>}(z)}{dz}\right|_{z=d_{2}} 
-	\epsilon_2\left.\frac{d\phi_{1<}(z)}{dz}\right|_{z=d_{2}} =-\frac{\sigma_2(z; {\bf q},\omega)}{\epsilon_{0}}
\end{split}
\label{condition2}
\end{equation}
In this way, we obtain the coefficients in Eq.~(\ref{match2}). By making use of these results for the coefficients in addition to linear response theory for the induced charge density, i.e.,  $\sigma_{1}(z;{\bf q},\omega)= \delta(z-z_1)e^{2} \chi^{(1)}({\bf q},\omega) \phi_{1<}(z=d_1)$ at the second monolayer, we obtain  

\begin{eqnarray}
\left(
\begin{array}{cc}
a_1\\\\
b_1
\end{array}\right)&=& 
\frac{1}{2\epsilon_{2}}\left(
\begin{array}{cc}
 \left(\epsilon_{2} +\frac{e^2}{\epsilon_{0}q}\chi^{(1)}({\bf  q},\omega)\right) e^{qd_1}  & e^{qd_1} \\\\
\left(\epsilon_{2} -\frac{e^2}{\epsilon_{0}q}\chi^{(1)}({\bf  q},\omega)\right) e^{-qd_1}  & -e^{-qd_1} \\
\end{array}
\right)
\left(
\begin{array}{cc}
e^{-qd_1} 
 & e^{qd_1} \\\\
  \epsilon_1 e^{-qd_1} 
 & - \epsilon_1  e^{qd_1}  
\end{array}
\right)
 \left(
\begin{array}{cc}
a_0\\\\
b_0
\end{array}\right).
\label{GAG2}
\end{eqnarray}
Combining Eqs.~(\ref{START}) and (\ref{GAG2}), we obtain the following explicit results for $a_1$ and $b_1$:

\begin{equation}
\begin{split}
a_1=   &\frac{1}{4\epsilon_{1}\epsilon_{2}}\left[(\epsilon_{1}+\frac{e^{2}\chi^{(0)}}{\epsilon_{0}q})(1-g)+(1+g)\right]\left(\epsilon_{2}+\frac{e^{2}\chi^{(1)}}{\epsilon_{0}q}+\epsilon_{1}\right)\\&+\frac{1}{4\epsilon_{1}\epsilon_{2}}\left[(\epsilon_{1}-\frac{e^{2}\chi^{(0)}}{\epsilon_{0}q})(1-g)-(1+g)\right]\left(\epsilon_{2}+\frac{e^{2}\chi^{(1)}}{\epsilon_{0}q}-\epsilon_{1}\right)e^{2qd_{1}}\\
b_1=  &\frac{1}{4\epsilon_{1}\epsilon_{2}}\left[(\epsilon_{1}+\frac{e^{2}\chi^{(0)}}{\epsilon_{0}q})(1-g)+(1+g)\right]\left(\epsilon_{2}-\frac{e^{2}\chi^{(1)}}{\epsilon_{0}q}-\epsilon_{1}\right)e^{-2qd_{1}}\\
&+\frac{1}{4\epsilon_{1}\epsilon_{2}}\left[(\epsilon_{1}-\frac{e^{2}\chi^{(0)}}{\epsilon_{0}q})(1-g)-(1+g)\right]\left(\epsilon_{2}-\frac{e^{2}\chi^{(1)}}{\epsilon_{0}q}+\epsilon_{1}\right)
\label{TL2}
\end{split}
\end{equation}

\medskip
\par
Finally, at the third monolayer, from the continuity and discontinuity conditions involving $\phi_{1<}(z)$ and $\phi_{1>}(z)$ evaluated at $z=d_2$, and using the linear response theory for the induced charge density $\sigma_{2}(z;{\bf q},\omega)= \delta(z-z_2)e^{2} \chi^{(2)}({\bf q},\omega) \phi_{1>}(z=d_2)$, we obtain

\begin{eqnarray}
\left(
\begin{array}{cc}
a_{1}\\
b_{1}
\end{array}
\right)
=
 \frac{1}{2\epsilon_2}
\left(
\begin{array}{cc}
\epsilon_2e^{qd_2}   & e^{qd_2} \\\\
\epsilon_2e^{-qd_2}   & -e^{-qd_2} 
\end{array}
\right)
\left(
\begin{array}{cc}
1 \\
1-\frac{e^2}{\epsilon_0q}\chi^{(2)}({\bf  q},\omega)
\end{array}
\right)  t_1e^{-qd_2}
\label{3L}
\end{eqnarray} 
This equation relates  $t_1$ to $a_1$, and $b_1$. Eliminating $t_1$ from these two equations, we obtain

\begin{equation}
\frac{b_1}{a_1}=
\frac{\epsilon_2-1+\frac{e^2}{\epsilon_0q}\chi^{(2)}({\bf q},\omega)}{ \epsilon_2+1-\frac{e^2}{\epsilon_0q}\chi^{(2)}({\bf q},\omega)}
\  \  e^{-2qd_2} \  .
\label{GFIN}
\end{equation}

Combining Eqs. (\ref{TL2})  and  (\ref{GFIN}), the resulting equation we obtain for the SRF of three layered structure in vacuum as depicted in Fig.~\ref{three-layer} is

\begin{eqnarray}
g_{3}({\bf q},\omega)&=\frac{N_{3}({\bf q},\omega)}{D_{3}({\bf q},\omega)}
\end{eqnarray}

\begin{equation}
\begin{split}
N_{3}({\bf q},\omega)=&\frac{1}{4\epsilon_{1}\epsilon_{2}}\left[\epsilon_{2}+1-\frac{e^{2}\chi^{(2)}}{\epsilon_{0}q}\right]\\
&\times\left\{e^{-2qd_{1}}\left(\epsilon_{2}-\frac{e^{2}\chi^{(1)}}{\epsilon_{0}q}-\epsilon_{1}\right)\left(\epsilon_{1}+\frac{e^{2}\chi^{(0)}}{\epsilon_{0}q}+1\right)+\left(\epsilon_{2}-\frac{e^{2}\chi^{(1)}}{\epsilon_{0}q}+\epsilon_{1}\right)\left(\epsilon_{1}-\frac{e^{2}\chi^{(0)}}{\epsilon_{0}q}-1\right)\right\}\\
&-\frac{1}{4\epsilon_{1}\epsilon_{2}}\left[\epsilon_{2}-1+\frac{e^{2}\chi^{(2)}}{\epsilon_{0}q}\right]e^{-2qd_{2}}\\
&\times\left\{\left(\epsilon_{2}+\frac{e^{2}\chi^{(1)}}{\epsilon_{0}q}+\epsilon_{1}\right)\left(\epsilon_{1}+\frac{e^{2}\chi^{(0)}}{\epsilon_{0}q}+1\right)+e^{2qd_{1}}\left(\epsilon_{2}+\frac{e^{2}\chi^{(1)}}{\epsilon_{0}q}-\epsilon_{1}\right)\left(\epsilon_{1}-\frac{e^{2}\chi^{(0)}}{\epsilon_{0}q}-1\right)\right\}
\label{N3}
\end{split}
\end{equation}

\begin{equation}
\begin{split}
D_{3}({\bf q},\omega)=&\frac{1}{4\epsilon_{1}\epsilon_{2}}\left[\epsilon_{2}-1+\frac{e^{2}\chi^{(2)}}{\epsilon_{0}q}\right]e^{-2qd_{2}}\\
&\times\left\{\left(\epsilon_{2}+\frac{e^{2}\chi^{(1)}}{\epsilon_{0}q}+\epsilon_{1}\right)\left(-\epsilon_{1}-\frac{e^{2}\chi^{(0)}}{\epsilon_{0}q}+1\right)+e^{2qd_{1}}\left(\epsilon_{2}+\frac{e^{2}\chi^{(1)}}{\epsilon_{0}q}-\epsilon_{1}\right)\left(-\epsilon_{1}+\frac{e^{2}\chi^{(0)}}{\epsilon_{0}q}-1\right)\right\}\\
&-\frac{1}{4\epsilon_{1}\epsilon_{2}}\left[\epsilon_{2}+1-\frac{e^{2}\chi^{(2)}}{\epsilon_{0}q}\right]\\
&\times\left\{e^{-2qd_{1}}\left(\epsilon_{2}-\frac{e^{2}\chi^{(1)}}{\epsilon_{0}q}-\epsilon_{1}\right)\left(-\epsilon_{1}-\frac{e^{2}\chi^{(0)}}{\epsilon_{0}q}+1\right)+\left(\epsilon_{2}-\frac{e^{2}\chi^{(1)}}{\epsilon_{0}q}+\epsilon_{1}\right)\left(-\epsilon_{1}+\frac{e^{2}\chi^{(0)}}{\epsilon_{0}q}-1\right)\right\}
\label{D3_}
\end{split}
\end{equation}
\medskip
\par

From the SRF for three layers, we can deduce the SRFs for different structures. So, 
we now consider the case when three monolayers are embedded in a homogeneous dielectric medium. We can deduce the SRF for this case by setting $\epsilon_{1}=\epsilon_{2}=1.0$ into $N_{3}({\bf q},\omega)$ and $D_{3}({\bf q},\omega)$ in Eq.~(\ref{D3_}). Subsequently, with the replacement $e^{2}\Rightarrow\frac{e^{2}}{\epsilon_{b}}$, where $\epsilon_{b}$ is the dielectric constant of the homogeneous dielectric medium, setting $k=\frac{1}{4\pi\epsilon_{0}}$, and factoring out $2$, we obtain the dielectric function for three layers in a uniform background as

\begin{equation}
\begin{aligned}[b]
\epsilon_{3}({\bf q},\omega)=\\
(1-&k\frac{2\pi e^{2}}{\epsilon_{b}q}\chi^{(0)}({\bf q},\omega))\biggr[\left(1-k\frac{2\pi e^{2}}{\epsilon_{b}q}\chi^{(1)}({\bf q},\omega)\right)\left(1-k\frac{2\pi e^{2}}{\epsilon_{b}q}\chi^{(2)}({\bf q},\omega)\right)-(k\frac{2\pi e^{2}}{\epsilon_{b} q})^{2}e^{-2q(d_{2}-d_{1})}\chi^{(1)}({\bf q},\omega)\chi^{(2)}({\bf q},\omega)\biggr]\\
&+k\frac{2\pi e^{2}}{\epsilon_{b}q}\chi^{(0)}({\bf q},\omega)\biggr[-k\frac{2\pi e^{2}}{\epsilon_{b}q}\chi^{(1)}({\bf q},\omega)\left(1-k\frac{2\pi e^{2}}{\epsilon_{b}q}\chi^{(2)}({\bf q},\omega)\right)e^{-2qd_{1}}-(k\frac{2\pi e^{2}}{\epsilon_{b}q})^{2}e^{-2qd_{2}}\chi^{(1)}({\bf q},\omega)\chi^{(2)}({\bf q},\omega)\biggr]\\
&-(k\frac{2\pi e^{2}}{\epsilon_{b}q})^{2}\chi^{(0)}({\bf q},\omega)\chi^{(2)}({\bf q},\omega)e^{-2qd_{2}}
\label{3vacuum}
\end{aligned}
\end{equation}

We emphasize that this expression assumes that with the assistance of the dielectric media between layers, the electrons are confined to the monolayers and there is no hopping between them. Additionally, we neglected any hybridization of the polarization functions in different monolayers.
\medskip
\par
At this point, we wish to make it clear that the dispersion function for the collective modes for three monolayers embedded in a dielectric medium was determined in the paper by Das Sarma and Madhukar\cite{Sarma} and presented in Eq.~(11) of their paper. This was written as $|\epsilon|=|\delta_{ij}-V_{ij}\chi_{jj}|$. In our notation, $\chi_{00}({\bf q},\omega), \chi_{11}({\bf q},\omega)$ and $\chi_{22}({\bf q},\omega)$ are $\chi^{(0)}({\bf q},\omega), \chi^{(1)}({\bf q},\omega)$ and $\chi^{(2)}({\bf q},\omega)$ which are the noninteracting polarizability in each layer. After a little algebra, it can be shown that Eq.~(\ref{3vacuum}) can be expressed as follows, in full agreement with Ref. \cite{Sarma} in SI unit. We have

\begin{eqnarray}
\epsilon_{3}({\bf q},\omega)=
\begin{vmatrix}
1-V_{00}\chi^{(0)}(\bf{q},\omega) &  -V_{01}\chi^{(0)}(\bf{q},\omega) &  -V_{02}\chi^{(0)}(\bf{q},\omega) 
\\\\ 1-V_{10}\chi^{(1)}(\bf{q},\omega) &  -V_{11}\chi^{(1)}(\bf{q},\omega) &  -V_{12}\chi^{(1)}(\bf{q},\omega) 
\\\\ 1-V_{20}\chi^{(2)}(\bf{q},\omega) &  -V_{21}\chi^{(2)}(\bf{q},\omega) &  -V_{22}\chi^{(2)}(\bf{q},\omega) 
\end{vmatrix}
\label{3by3}
\end{eqnarray}
Here, $V_{00} = V_{11} = V_{22}= k\frac{2\pi e^{2}}{\epsilon_{b} q}$ are the magnitude of Coulomb interaction of charge carriers in each monolayer. Here,  $\epsilon_{b}$ is the average background dielectric constant. $V_{01}=V_{10}=k\frac{2\pi e^{2}}{\epsilon_{b} q}e^{-qd_{1}}$ are the Coulomb interactions between the first and second monolayers which are exponentially dependent on the separation between the first and second layer, $d_{1}$. Accordingly, $V_{02}=V_{20}=k\frac{2\pi e^{2}}{\epsilon_{b} q}e^{-qd_{2}}$ are the Coulomb interactions between the first and third layers, and $V_{12}=V_{21}=k\frac{2\pi e^{2}}{\epsilon_{b}  q}e^{-q(d_{2}-d_{1})}$ are the Coulomb interactions between the second and third layers. It is important to make it clear that when analyzing the determinantal form, that we keep in mind that it is the same expression as Eq.~(\ref{3vacuum}), which was derived from the denominator of our SRF for three monolayers. Therefore, this validates the validity of our SRF approach for three monolayers.

\begin{figure}[h]
\captionsetup{singlelinecheck=off,justification=raggedright}
\centering
\includegraphics[width=.47\columnwidth]{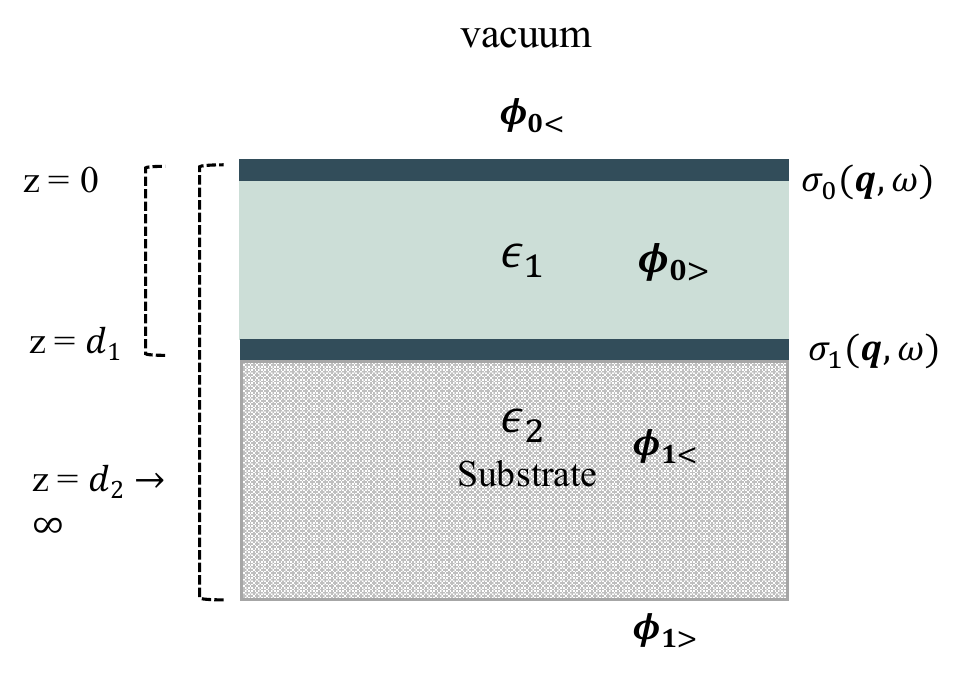}
\caption{(Color online) Schematic representation of two electron monolayers on a thick substrate with dielectric constant $\epsilon_{2}$. The separation between the pair is arbitrary. The induced surface charge densities are $\sigma_{0}({\bf q},\omega)$  and $\sigma_{1}({\bf q},\omega)$  located  at z= $0$ and  z=$d_{1}$, respectively. The layers are separated by material with dielectric constant $\epsilon_{1}$.}
\label{2L_sub}
\end{figure}

\subsection{Two layered structure placed on a substrate and exposed to vacuum above}

We now keep only the two layers located at $z=0$ and $z=d_1$ by setting $\chi^{(2)}=0$ and taking the limit $d_{2}\rightarrow\infty$, $\epsilon_{1}\neq\epsilon_{2}\neq 1$ from the SRF for three layers. In this way, we arrive at the SRF for a pair of 2D layers where the first layer sits on the surface with vacuum on one side and a slab with dielectric constant $\epsilon_{1}$ on the other, whereas the second layer is embedded in media with dielectric constants $\epsilon_{1}$ and $\epsilon_{2}$ as depicted in Fig.~\ref{2L_sub}. We have

\begin{eqnarray}
g_{2s}({\bf q},\omega)&=\frac{N_{2s}({\bf q},\omega)}{D_{2s}({\bf q},\omega)}
\label{g2s}
\end{eqnarray}

\begin{equation}
\begin{split}
N_{2s}({\bf q},\omega)=& \frac{1}{4\epsilon_{1}\epsilon_{2}}\left[\epsilon_{2}+1\right]\\
&\times \left\{ e^{-2qd_{1}}\left(\epsilon_{2}-\frac{e^{2}\chi^{(1)}}{\epsilon_{0}q}-\epsilon_{1}\right)\left(\epsilon_{1}+\frac{e^{2}\chi^{(0)}}{\epsilon_{0}q}+1\right)+\left(\epsilon_{2}-\frac{e^{2}\chi^{(1)}}{\epsilon_{0}q}+\epsilon_{1}\right)\left(\epsilon_{1}-\frac{e^{2}\chi^{(0)}}{\epsilon_{0}q}-1\right)    \right\}
\end{split}
\end{equation}

\begin{equation}
\begin{split}
D_{2s}({\bf q},\omega)=&-\frac{1}{4\epsilon_{1}\epsilon_{2}}\left[\epsilon_{2}+1\right]\\
&\times \left\{e^{-2qd_{1}}\left(\epsilon_{2}-\frac{e^{2}\chi^{(1)}}{\epsilon_{0}q}-\epsilon_{1}\right)\left(-\epsilon_{1}-\frac{e^{2}\chi^{(0)}}{\epsilon_{0}q}+1\right)+\left(\epsilon_{2}-\frac{e^{2}\chi^{(1)}}{\epsilon_{0}q}+\epsilon_{1}\right)\left(-\epsilon_{1}+\frac{e^{2}\chi^{(0)}}{\epsilon_{0}q}-1\right)\right\}.
\label{D2s}
\end{split}
\end{equation}
Additionally, by setting, $\epsilon_{1}=\epsilon_{2}=1$, we derive the SRF for a pair of  2D layers {\em in vacuo\/}. But, if these layers are embedded in a medium with uniform background dielectric constant $\epsilon_b$, we make the replacement  $e^{2}\Rightarrow e^{2}/\epsilon_{b}$ to take account of dielectric screening, we obtain from $D_{2s}({\bf  q},\omega)$ the well known formula for the plasmon dispersion function for a pair of layers in a uniform background medium also shown in Ref.~\cite{Sarma}, which is

\begin{equation}
\epsilon_{2}({\bf q},\omega)=1-k\frac{2\pi e^{2}}{q\epsilon_{b}}(\chi^{(0)} ({\bf q},\omega)
+\chi^{(1)} ({\bf q},\omega)
)+ \left(k\frac{2\pi e^{2}}{q\epsilon_{b}}\right)^{2}\left(1-e^{-2qd_{1}}\right)\chi^{(0)} ({\bf q},\omega)
\chi^{(1)} ({\bf q},\omega) \ .
\label{sarma}
\end{equation}

\medskip
\par

\subsection{Two layered structure freely suspended in vacuum}

It is interesting to consider a pair of 2D electron layers located at $z=0$ and $z=d_1$ with a dielectric $\epsilon_1$ in the space $0<z<d_1$ and vacuum in $z<0$ and $z>d_1$. We can derive the SRF for this two-layer structures freely suspended in vacuum directly from the boundary conditions. Equation.~(\ref{START}) still applies for the relationship between the coefficients  $a_0$ and $b_0$ for the electrostatic  potential  in the range $0\leq  z\leq  d_1$. Furthermore, when we apply the matching and discontinuity boundary conditions at the second  monolayer, we obtain the exact same form for the ratio $b_0/a_0$ like in the case presented in Eq.~(\ref{GFIN}) for $b_1/a_1$ for three layers except that the following replacements are obtained, i.e., $d_2\Rightarrow  d_1$, $\epsilon_{2}\Rightarrow \epsilon_{1}$ and $\chi^{(2)}({\bf q},\omega) \Rightarrow  \chi^{(1)}({\bf q},\omega)$. That is

\begin{equation}
\frac{b_0}{a_0}=
\frac{\epsilon_{1}-1+\frac{e^2}{\epsilon_0q}\chi^{(1)}({\bf q},\omega)}{ \epsilon_1+1-\frac{e^2}{\epsilon_0q}\chi^{(1)}({\bf q},\omega)}
\  \  e^{-2qd_1} \  .
\label{GFIN2}
\end{equation}

\begin{figure}[h]
\captionsetup{singlelinecheck=off,justification=raggedright}
\centering
\includegraphics[width=.5\columnwidth]{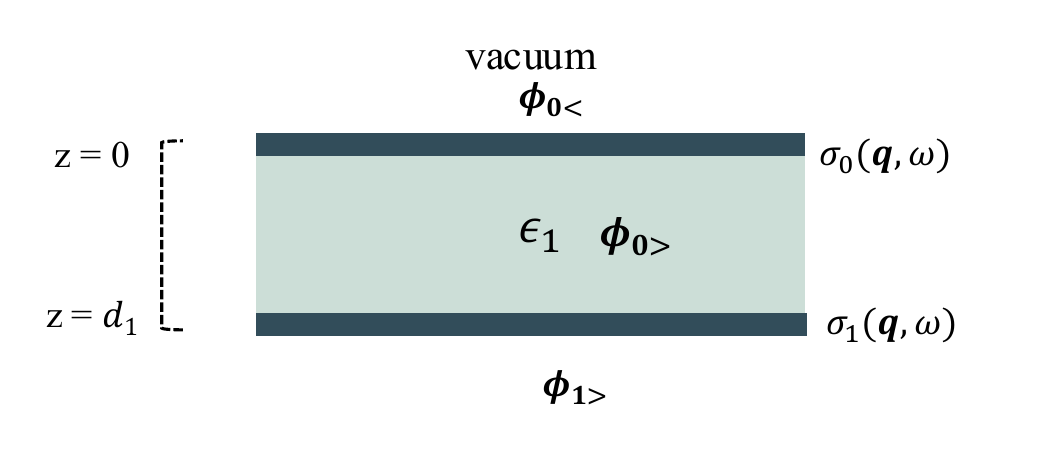}
\caption{(Color online)Schematic illustration of two freely suspended electron monolayers in vacuum with arbitrary separation  between them. The induced surface charge densities are $\sigma_{0}({\bf q},\omega)$ and $\sigma_{1}({\bf q},\omega)$ at z= $0$, z=$d_{1}$, respectively.  The layers are separated by a material with dielectric constant $\epsilon_{1}$.}
\label{two_layer}
\end{figure}

Combining Eqs.~(\ref {fracb0a0}) and (\ref{GFIN2}), the resulting equation which we obtain for the SRF for two layers in vacuum as depicted in Fig.~\ref{two_layer} is

\begin{eqnarray}
g_{2}({\bf q},\omega)&=\frac{N_{2}({\bf q},\omega)}{D_{2}({\bf q},\omega)}
\label{N2D2}
\end{eqnarray}

\begin{eqnarray}
N_{2}({\bf q},\omega)=-(1+\epsilon_{1}-\frac{e^{2}\chi^{(1)}}{q\epsilon_{0}})(1-\epsilon_{1}+\frac{e^{2}\chi^{(0)}}{q\epsilon_{0}})+(-1-\epsilon_{1}-\frac{e^{2}\chi^{(0)}}{q\epsilon_{0}})(\epsilon_{1}-1+\frac{e^{2}\chi^{(1)}}{q\epsilon_{0}})e^{-2qd_{1}}
\end{eqnarray}

\begin{eqnarray}
D_{2}({\bf q},\omega)=(1+\epsilon_{1}-\frac{e^{2}\chi^{(1)}}{q\epsilon_{0}})(1+\epsilon_{1}-\frac{e^{2}\chi^{(0)}}{q\epsilon_{0}})+(\epsilon_{1}-1+\frac{e^{2}\chi^{(1)}}{q\epsilon_{0}})(1-\epsilon_{1}-\frac{e^{2}\chi^{(0)}}{q\epsilon_{0}})e^{-2qd_{1}}
\label{2d}
\end{eqnarray}
\medskip
\par

We now compare the dispersion functions (a) when the two layers are embedded in a uniform dielectric background, with (b) the dispersion function when two parallel layers have space between them filled with the same dielectric and this structure is suspended freely in vacuum. We expect that dielectric screening, inter-layer separation, energy band structure and temperature will play crucial roles in the plasmon dispersions and their damping and for this reason we now further examine in detail this more general case. Interestingly, when the denominator of the SRF for two layers coating a dielectric medium, we multiply Eq.~(\ref{2d})   by a factor of $\frac{1}{4}$, and $e^{2}$ is replaced with $e^{2}/\epsilon_{b}$  for further illustration.    Consequently, the corresponding plasmon dispersion function obtained from the SRF for this pair of layers suspended in vacuum and separated by a dielectric can be expressed as

\begin{equation}
\begin{split}
\frac{1}{4}D_2({\bf q},\omega)=&\underbrace{\frac{1}{4}\left[(\epsilon_1+1)^2-(\epsilon_1-1)^2 e^{-2qd_1}\right]+\frac{1}{4}(1-\epsilon_1)(1+e^{-2qd_{1}})k\frac{2\pi e^2}{\epsilon_{b}q}\left(\chi^0({\bf  q},\omega)+\chi^1({\bf  q},\omega)\right)-1}_{\text{due to inhomogeneity}}\\
&+\begin{vmatrix}
\epsilon_{11}(\bf{q},\omega) & \epsilon_{12}(\bf{q},\omega) \\ \epsilon_{21}(\bf{q},\omega) & \epsilon_{22}(\bf{q},\omega)
\end{vmatrix}
\label{twoby2}
\end{split}
\end{equation}
with

\begin{eqnarray}
\epsilon_{11}({\bf q},\omega)&=&1- k\frac{2\pi e^2}{\epsilon_{b}q}\chi^0({\bf  q},\omega)
\nonumber\\
\epsilon_{22}({\bf q},\omega)&=&1- k\frac{2\pi e^2}{\epsilon_{b}q}\chi^1({\bf  q},\omega)
\nonumber\\
\epsilon_{12}({\bf q},\omega)&=& k\frac{2\pi e^2}{\epsilon_{b}q}\chi^0({\bf  q},\omega)e^{-qd_1}
\nonumber\\
\epsilon_{21}({ \bf q},\omega)&=& k\frac{2\pi e^2}{\epsilon_{b}q}\chi^1({\bf  q},\omega)e^{-qd_1}\  
\label{4eqns}
\end{eqnarray}
which is such that when we set $\epsilon_1=1$, only the determinental term survives as expected for the dispersion equation for a pair of 2D layers with a uniform background as it is well known from Ref. \cite{Sarma}. The effects due to the additional terms in Eq.~(\ref{twoby2}) will be investigated from our results by carrying out numerical  calculations in the presence and absence of these contributions to the dispersion equation. Although there are always two branches for a pair of 2D layers, these extra terms can affect the plasmon excitation energies, the damping of the allowed acoustic and optical modes and their separation as well as their anti-crossing behavior.

\medskip
\par

As we noted above, the determinant term which appears in Eq.~(\ref{twoby2}) is the dielectric function for a standard double layered structure embedded in a medium with a uniform dielectric constant. This result is well known from Ref. \cite{Sarma} and is given in Eq.~(\ref{sarma}). Therefore, we have formally shown that the SRF for a coated structure has extra terms in addition to the expression for the dielectric function of the standard double-layered structure with a uniform dielectric background. We further investigate the difference between plasmonic properties of this standard double-layer geometry and the coated structure placed on a substrate by plotting the plasmon dispersions in  Sec~\ref{sec4}.


\subsection{Single layer placed on a thick substrate}
\begin{figure}[h]
\captionsetup{singlelinecheck=false, justification=raggedright}
\centering
\includegraphics[width=.45\columnwidth]{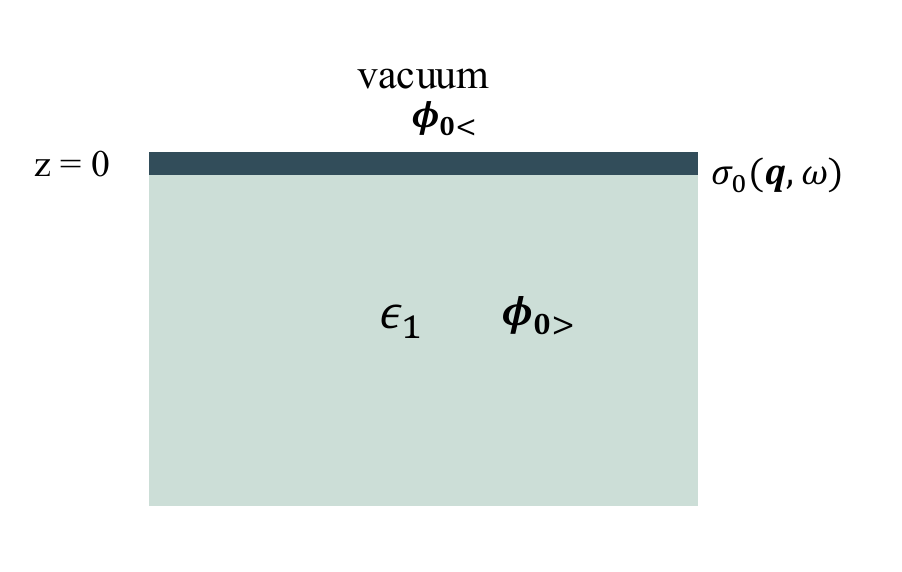}
\caption{(Color online) Schematic illustration of a monolayer placed on a thick substrate with dielectric constant $\epsilon_{1}$. There is vacuum above the structure. The induced surface charge density is $\sigma_{0}(q,\omega)$ at z= $0$. }
\label{sub1}
\end{figure}

From Eq.~(\ref{N2D2}), we deduce the SRF for a single layer on a substrate by setting $\chi^{(1)}({\bf q},\omega)=0$, the second dielectric constant $\epsilon_{2}=1$, and also taking the limit of $d_{1}\to\infty$. Factoring out the common quantity $(1+\epsilon_{0})$, we arrive at

\begin{equation}
g_{1s}({\bf q},\omega)=\frac{N_{1s}({\bf   q},\omega)}{D_{1s}({\bf q}
,\omega)}
\label{1g1}
\end{equation}

\begin{equation}
N_{1s}({\bf q},\omega)= -\left(1-\epsilon_{1}+\frac{e^{2}\chi^{(0)}({\bf q},\omega)}{q\epsilon_{0}}\right)
\label{N1}
\end{equation}

\begin{equation}
D_{1s}({\bf q},\omega)= 1+\epsilon_{1}-\frac{e^{2}\chi^{(0)}({\bf q},\omega)}{q\epsilon_{0}}
\label{D1}
\end{equation}
On simplification, we obtain the following equation, which agrees with that in Ref. \cite{Gumbs} and Ref. \cite{BNJP}.

\begin{equation}
 g_{1s}(\mathbf{q},\omega)=1-\frac{2}{1+\epsilon_{1}-\frac{e^{2}\chi^{(0)}({\bf q},\omega)}{q\epsilon_{0}}}
=1-\frac{1}{\bar{\epsilon_b}\epsilon({\bf q},\omega)}
\label{Persson1}
\end{equation}
where the dielectric function is 

\begin{equation}
\epsilon({\mathbf{q}},\omega)= 1-\frac{2\pi e^2}{\bar{\epsilon_{s}}q} \chi_{0}({\bf q},\omega)\  .
\label{Persson2}
\end{equation}
which can be also directly derived from the zero of Eq.~(\ref{D1}), the denominator of the SRF. Here, $\bar{\epsilon_s}=4\pi\epsilon_0\bar{\epsilon_b}$, and $\bar{\epsilon_b}$ is the average background dielectric constant $(1+\epsilon_{1})/2$.
\medskip
\par

We carried out detailed analytical calculations for one, two and three layers and showed that there are obvious quantitative differences between the dispersion equations for a homogeneous versus inhomogeneous dielectric background. From the poles of our SRF, the plasmon dispersion relation can be derived since denominator of SRF, $D(\mathbf{q},\omega)$, is proportional to the original expression of dielectric function. As a matter of fact, in the paper by Horing, et al. \cite{Horing}, it was shown that  when a beam of fast moving charged particles interact with a 2D layer, the rate of loss of energy is determined by the loss function $-\Im(1/\epsilon({\bf q},\omega))$. Therefore, the advantage of this method is that the SRF can help determine the plasmon excitation relations, loss function, and absorption coefficient for conditions such as multilayer structures separated by dielectric media and placed on a substrate and exposed to vacuum above. Also, this SRF can be applied to any 2D electron material which will serve as the protective layer of the dielectric material. Moreover, one may separate the contributions to the loss function from particle-hole modes and plasmon excitations. Now, we want to put these results to the test with a specific material. Our choice of material is gapped TSDM due to its interesting energy  bands which are tilted and half-linear in one direction and half-parabolic in the perpendicular direction of wave vector space. Also, interest in this material is rapidly growing among both experimentalists and theoreticians due to its potential for applications in optoelectronics and photonics. In the next section, we will use these SRFs to obtain plasmon dispersion, loss function, and absorption coefficient for the layers of gapped TSDMs which are anisotropic in nature placed in the arbitrary dielectric background.
\medskip
\par


\section{Polarization function and plasma excitations}
\label{sec4}

\subsection{Plasmons for Single layer on a thick substrate}

 We employ the density matrix $\hat{\rho}({\bf  r},t)$ to determine the charge density fluctuation $\sigma({\bf r},\omega)$. Starting with the non-dissipative equation of motion \(  i\hbar\partial\hat{\rho}({\bf  r},t)/\partial  t=[\hat{H}(t),\hat{\rho}({\bf  r},t)]\), we separate both the Hamiltonian $ \hat{H}(t)= \hat{H}_0+\hat{H}_1(t)$ and the density matrix $ \hat{\rho}(t)= \hat{\rho}_0+\hat{\rho}_1(t) $ into an unperturbed time-independent part and a time-dependent term. Retaining only linear terms, we obtain $\sigma({\bf r},\omega)=e^2\int d{\bf r}^\prime\ \chi^{(0)}({\bf  r},{\bf r}^\prime,\omega)\phi({\bf  r}^\prime)$, where $\phi({\bf  r} )$ is the induced electrostatic potential, and the polarization is given by

\begin{equation}
\begin{split}
&\chi^{(0)}(\mathbf{r},\mathbf{r}^\prime,\omega)= \sum_{j,j}\frac{ f_{0}(\varepsilon_{i})-f_{0}(\varepsilon_{j})  }{\hbar(\omega+i\delta^+)-(\varepsilon_{j}-\varepsilon_{i})}\Psi_{j}^{\dag}(\mathbf{r }^\prime)\Psi_{i}(\mathbf{r}^\prime)\Psi_{i}^\dag(\mathbf{r}) 
\Psi_{j}(\mathbf{r})\  ,
\label{chi01}
\end{split}
\end{equation}

where $\Psi_{i}(\mathbf{r})$ are eigenfunctions of the unperturbed Hamiltonian and the indices $i,j$ label these eigenstates. If we now set $\Psi_{j}(\mathbf{r}) =\ket{\mathbf{k},\nu}\exp(i\mathbf{k} \cdot \mathbf{r} )/\sqrt{A} $, where $A$ is a normalization area, then Fourier transforming Eq.\  (\ref{chi01}), we obtain 
\begin{equation}
\chi^{(0)}(\mathbf{q},\omega)=2\sum_{\nu,\nu^\prime}\int \  \frac{d{\bf  k}}{(2\pi)^2} \frac{ f_0\left(\upsilon_{{\bf  k},\nu}\right) -      f_0\left(\upsilon_{{\bf  k}+{\bf q},\nu^\prime}\right) }{\hbar\omega+i\delta^+ -(\epsilon_{{\bf  k}+{\bf q},\nu^\prime}  -   \epsilon_{{\bf  k}  ,\nu } )}   {\cal F}({\bf  k},{\bf q};\lambda,\lambda^\prime)      \  ,
\label{chi02}
\end{equation}
where the factor of  $2$ is introduced to account for spin, degeneracy and the overlap function is given by 
\begin{equation}   
{\cal F}({\bf  k},{\bf q};\lambda\lambda^\prime)  \equiv |\left< {\bf  k} +{\bf  q},\lambda^\prime | {\bf  k},\lambda  \right>|^2\  .
\end{equation}

We further investigate the plasmon dispersion in the long wavelength limit. The expression below corresponds to the overlap function for gapped TSDM when $\bf{q}=0$, and will be employed in a Taylor series expansion of the overlap function in the small $\bf{q}$ limit. We have

\begin{eqnarray}
|\bra{\mathbf{k},\lambda^\prime}\ket{\mathbf{k},\lambda}|^{2}&=&
\left\{  2+\frac{\Delta^{2}}{(\hbar\upsilon_{F}k_{y})^{2}+(\hbar ak_{x}^{2})^{2}}\right\}^{-2}
\nonumber\\
&\times&\left\{ 1+2\lambda\lambda^\prime\left(1+\frac{\Delta^{2}}{(\hbar\upsilon_{F}k_{y})^{2}+(\hbar ak_{x}^{2})^{2}}\right)+\left(1+\frac{\Delta^{2}}{(\hbar\upsilon_{F}k_{y})^{2}+(\hbar ak_{x}^{2})^{2}}  \right)^{2}      \right\}\   .
\label{TLO}
\end{eqnarray}
This overlap is equal to `one’, when $\lambda=\lambda^\prime$, as imposed by the normalization condition. Furthermore, we have for $\lambda\neq \lambda^\prime$
\begin{equation}
|\bra{\mathbf{k},\lambda^\prime=+}\ket{\mathbf{k},\lambda=-}|^{2}= 
\left\{ 1+ \frac{2}{\Delta^{2}} \left[(\hbar\upsilon_{F}k_{y})^{2}+(\hbar ak_{x}^{2})^{2}\right] \right\}^{-2}
\label{TLO2}
\end{equation}
\medskip
\par
When we neglect inter-valley scattering and include intra-band transitions $\lambda=\lambda^\prime=1$ only, the polarization function in long wavelength limit is given by the below expression. Here, we also assume that the damping is negligible such that $\delta^{+}\ll\omega$.  We obtain

\begin{equation}
\chi^{(0)}({\bf  q},\omega)\approx 
\frac{1}{(\hbar\omega)^2}
\frac{2}{(2\pi)^{2}} \int_{-\infty}^\infty dk_{x}\int_{-\infty}^\infty dk_{y}\  f(\varepsilon({\bf k},\nu)) \left\{ \alpha_{2x}(k_x,k_y)q_x^2 + \beta_{2y}(k_x,k_y)q_y^2  \right\}=\frac{P^{(0)}}{(\hbar\omega)^2}\  ,
\label{Aeqn}
\end{equation}
where the  coefficients are defined as

\begin{equation}
\alpha_{2x}(k_x,k_y) \equiv     2C\frac{k_x^2}{{\cal  D}(k_x,k_y)}  \left\{ 3-    C\frac{2k_x^4}{{\cal  D}^2(k_x,k_y)  } \right\}\  , \  \  \  \
\beta_{2y}(k_x,k_y) \equiv   B\frac{1}{{\cal  D}(k_x,k_y)}  \left\{ 1- B\frac{k_y^2}{{\cal  D}^2(k_x,k_y)}   \right\}
\label{Beqn}
\end{equation}
with ${\cal D}(k_x,k_y)=\left\{Bk_y^2+Ck_x^4+\Delta^2 \right\}^{1/2}$ with $B=(\hbar\upsilon_F)^2$ and $C=(\hbar  a)^2$. We now  have an analytic expression for the plasmon dispersion relation in the long wavelength limit when only intraband transitions contribute and inter-valley scatterings do not play a role.  As a matter of fact, the plasmon dispersion relation for a monolayer on a substrate is given by the poles of the SRF in Eq.~(\ref{Persson2}). Combining it with  Eqs.~(\ref{Aeqn})  and (\ref{Beqn}),  we obtain the plasmon dispersion  relation in the $q_x,q_y$ plane as

\begin{equation}
(\hbar\omega_p)^2\approx\frac{2\pi e^2}{\bar{\epsilon_s} q}\left\{  \frac{2}{(2\pi)^{2}} \int dk_{x}\int dk_{y}\  f(\varepsilon({\bf k},\nu)) \left\{ \alpha_{2x}(k_x,k_y)q_x^2 + \beta_{2y}(k_x,k_y)q_y^2  \right\}  \right\}=\frac{2\pi e^2}{\bar{\epsilon_s} q}P^{(0)}\  . 
\label{qxom}
\end{equation}

\begin{figure}[h]
\centering
\begin{subfigure}{.4\textwidth}
\refstepcounter{subfigure}
\includegraphics[width=.80\columnwidth]{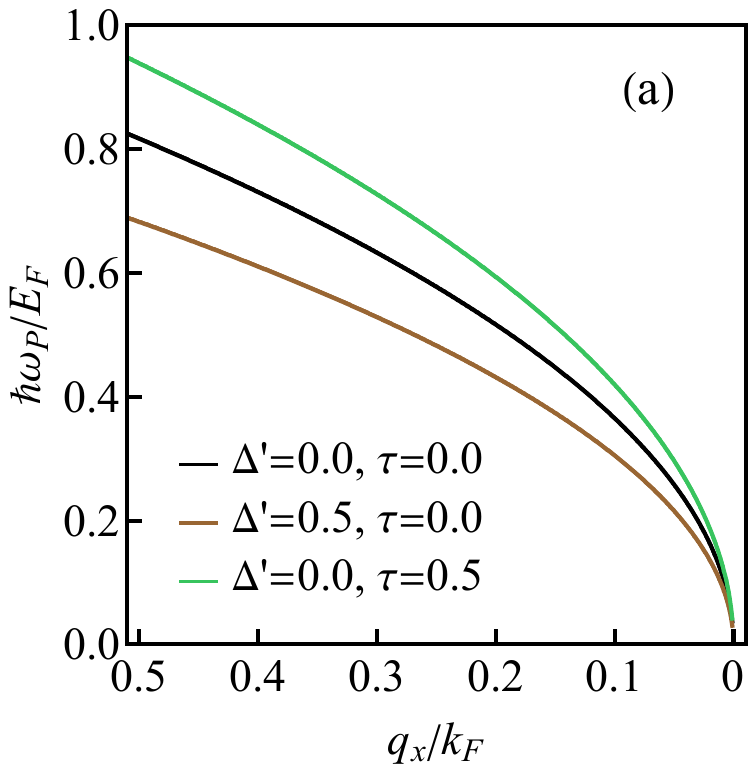}
\caption*{}
\label{qx} 
\end{subfigure}%
\hspace{0pt}
\begin{subfigure}{.4\textwidth}
\refstepcounter{subfigure}
\includegraphics[width=.80\columnwidth]{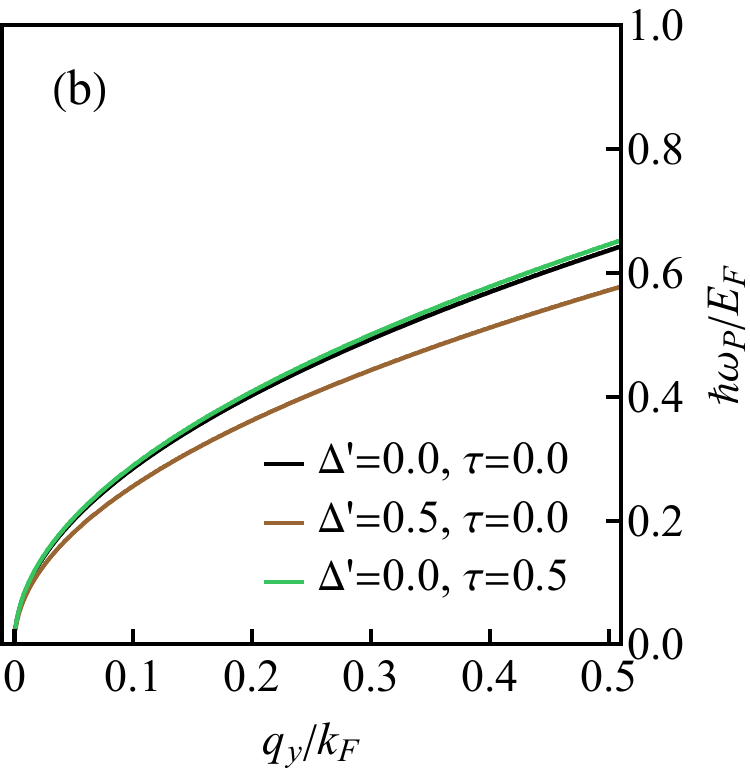}
\caption*{}
\label{qy}
\end{subfigure}\captionsetup{singlelinecheck=off,justification=raggedright}
\caption{(Color online) Plasmon dispersion in the long wavelength limit for a monolayer of semi-Dirac material covering a thick dielectric substrate with $\epsilon_{1}=5.0$. The frequency is in unit of $E_{F}/\hbar$ where $E_{F}=\hbar \upsilon_{F}k_{F}$ and we chose Fermi velocity  of graphene $v_{F}=10^{6}$m/s. Penal (a) is plotted as a function of $q_{x}/k_{F}$ when $q_{y}=0$. The black curve corresponds to a semi-Dirac layer without a gap and no tilting, the brown curve is for a layer with $\Delta'=0.5$, and the green curve shows the plasmon dispersion relation for a layer with tilting, $\tau=0.5$. For comparison, the $q_{x}/k_{F}$ increases from right to left.  (b) The plasmon frequency is plotted as a function of $q_{y}/k_{F}$ when $q_{x}  =0$.}
\label{expansion}
\end{figure}

\medskip
\par
We depict the numerical result of Eq.~(\ref{qxom}) in Fig.~\ref{expansion}, where the analysis is restricted to the regions where $q_{x}/k_{F}$ and $q_{y}/k_{F}$ are small. In this numerical calculations for the plasma dispersion of one layer of semi-Dirac material on a substrate, we scaled the wave vector and energy by $k_{F}$ and the cut-off Fermi energy of $E_F=\hbar\upsilon_F k_F$ at T=0 K. Since we assume that T=0 K, the Fermi-Dirac distribution function $f(\varepsilon({\bf k},\nu))$ becomes a Heaviside step function, $\Theta(E_F-\varepsilon(\mathbf{k},\nu))$ where the magnitude of the fermi velocity is $\upsilon_F=c/300=10^6 m/s$. It is known that in-plane dielectric constant of h-BN ranges from 3.29 to 3.76, while its out-of-plane dielectric constant ranges from 6.82 to 6.93 \cite{Akash}. We approximate the dielectric constant of the substrate as the mean value, $\epsilon_{1}=5.0$, which results in the average background dielectric constant of $\bar{\epsilon_{b}}=3.0$. We neglect Coulomb scattering of excited quasiparticle states and the subsequent effect on plasmon decay rates. However, we do take into account possible Landau damping by particle-hole modes, assuming that $\delta^{+}\ll\omega$ in the polarization function. In Fig.~\ref{qx}, the plasmon frequency is plotted with respect to $q_{x}/k_{F}$ when $q_{y} =0$, and it shows that when the energy gap parameter with magnitude $\Delta'=0.5$ is present and there is no tilting, the plasmon frequency is decreased. In contrast, when there is no gap but there is tilting with $\tau=0.5$, the plasmon frequency is increased. Figure.~\ref{qy} presents the plasmon frequency versus $q_{y}/k_{F}$ where $q_{x} =0$. In this case, when a gap $\Delta'$ is introduced, the plasmon frequency is decreased, and when tilting $\tau=0.5$ is present, the frequency is slightly increased. Overall, plots as functions of $q_{y}/k_{F}$ in Fig.~\ref{qy} are lower than for Fig.~\ref{qx} characterizing the anisotropic behavior in perpendicular momentum directions.

\begin{figure}[h]
\centering
\begin{subfigure}{.43\textwidth}
\refstepcounter{subfigure}
\includegraphics[width=1.\columnwidth]{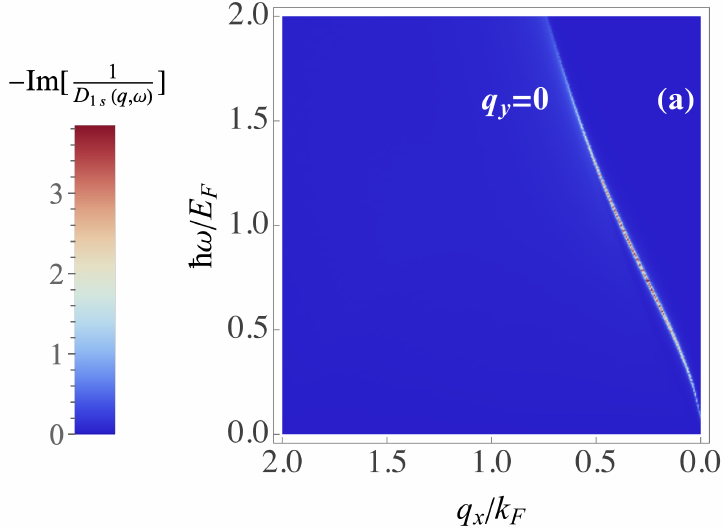}
\caption*{}
\label{density1qx_img} 
\end{subfigure}%
\hspace{20pt}
\begin{subfigure}{.43\textwidth}
\refstepcounter{subfigure}
\includegraphics[width=1.\columnwidth]{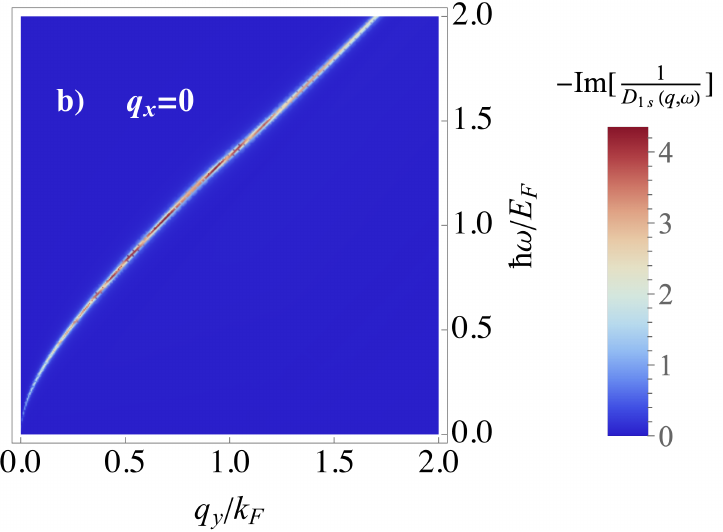}
\caption*{}
\label{density1qy_img}
\end{subfigure}\captionsetup{singlelinecheck=off,justification=raggedright}
\caption{(Color online) With $D_{1s}({\bf q},\omega)= 1+\epsilon_{1}-\frac{e^{2}\chi^{(0)}({\bf q},\omega)}{q\epsilon_{0}}$, we draw imaginary parts of $[-1/D_{1s}({\bf q},\omega)]$ for a monolayer of TSDM placed on a thick substrate with dielectric constant of $\epsilon_1=5.0$, where the bright line is a plasmon branch with frequency $\omega_P$. We chose tilting $\tau=0.1$, energy gap parameter $\Delta'=0.1$, and dephasing parameter $\hbar\delta^+/E_F=0.01$. (a) Frequency $\omega$ in unit of $E_{F}/\hbar$ is  plotted with respect to $q_{x}/k_{F}$ when $q_{y}=0$. (b) Frequency $\omega$ in unit of $E_F/\hbar$ is  plotted as a function of  $q_y/k_F$ when $q_x  =0$.}
\label{density1qxqy_re}
\end{figure}

\medskip
\par
Figure.~\ref{density1qxqy_re} is the density plot for imaginary parts of $-1/D_{1s}(q,\omega)$ which is proportional to the loss function of a monolayer with tilting $\tau=0.1$ and energy gap parameter $\Delta'=0.1$ placed on a thick substrate with dielectric constant of $\epsilon_{1}=5.0$. These results are based on the exact expression $D_{1s}$ given by Eq.~(\ref{D1}), and not the long wavelength approximation, but present density plots for $q_{x}/k_{F}$ and $q_{y}/k_{F}$ in the range of $0.0$ to $2.0$. Here we choose dephasing parameter $\hbar\delta^+/E_F=0.01$ and use Fermi velocity $\upsilon_{F}=10^{6}$ m/s. We observe bright lines which correspond to the plasmon branches with frequency $\omega_P$. The plasmon branches initially increase linearly as $q_{x}/k_{F}$ or $q_y/k_{F}$ is increased. The magnitude of plasmon frequency in the $q_{x} $ direction is slightly higher, indicating the anisotropic behavior similar to that shown in the Fig. \ref{expansion} obtained using a Taylor series expansion for the polarization function.

\begin{figure}[h]
\centering
\begin{subfigure}{.38\textwidth}
\refstepcounter{subfigure}
\includegraphics[width=0.9\columnwidth]{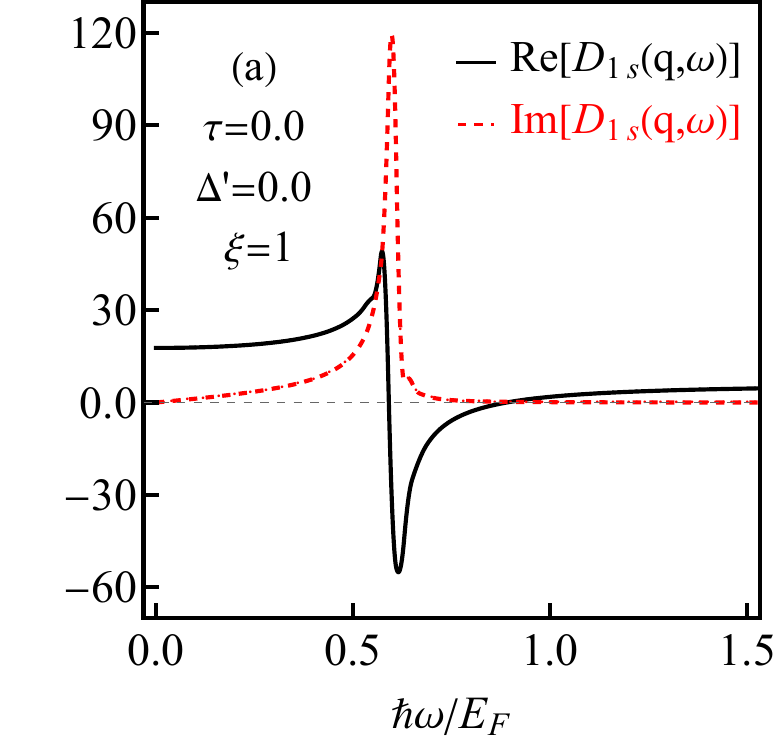}
\label{d_w1} 
\end{subfigure}
\hspace{0\textwidth}
\begin{subfigure}{.38\textwidth}
\refstepcounter{subfigure}
\includegraphics[width=0.9\columnwidth]{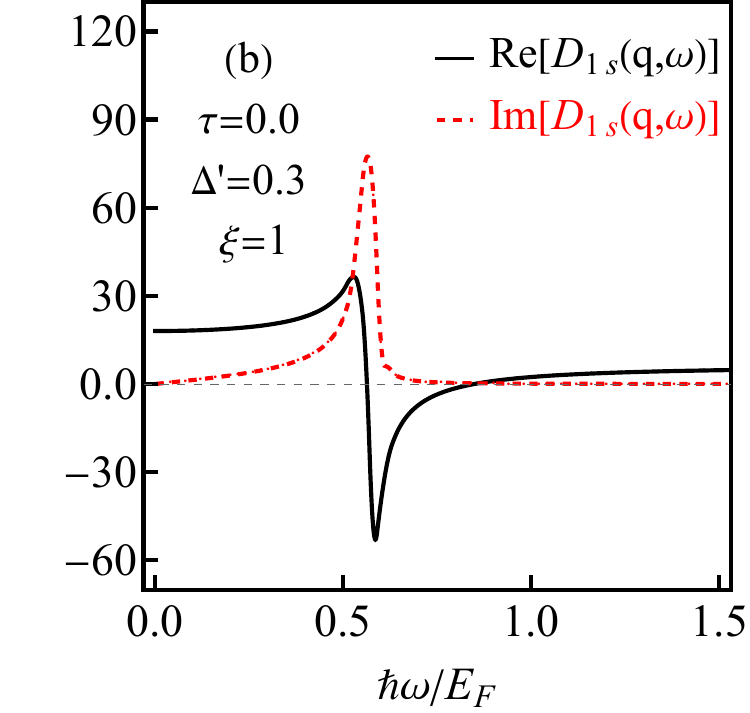}
\label{d_w2}
\end{subfigure}
\begin{subfigure}{.38\textwidth}
\refstepcounter{subfigure}
\includegraphics[width=0.9\columnwidth]{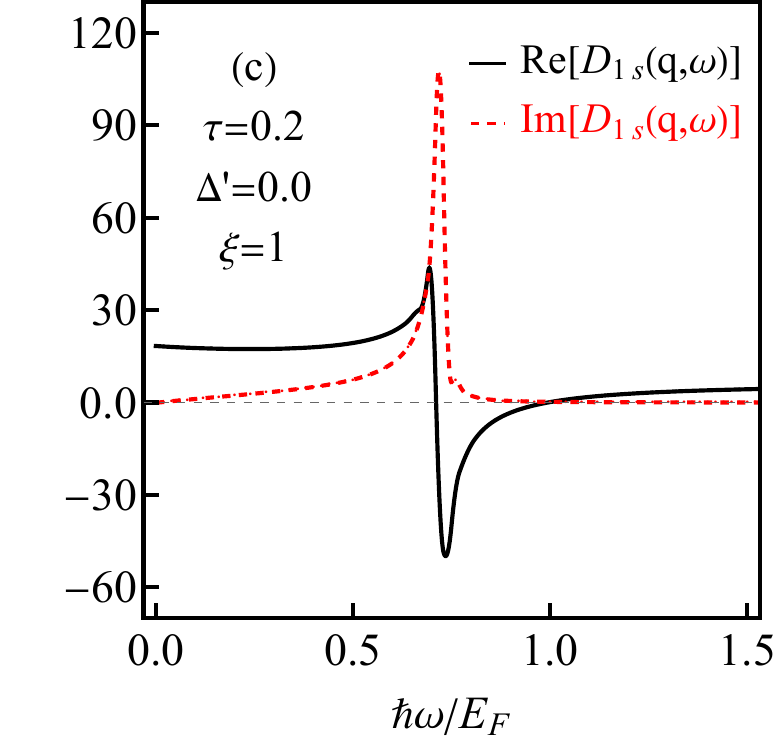}
\label{d_w3}
\end{subfigure}
\hspace{0\textwidth}
\begin{subfigure}{.38\textwidth}
\refstepcounter{subfigure}
\includegraphics[width=0.9\columnwidth]{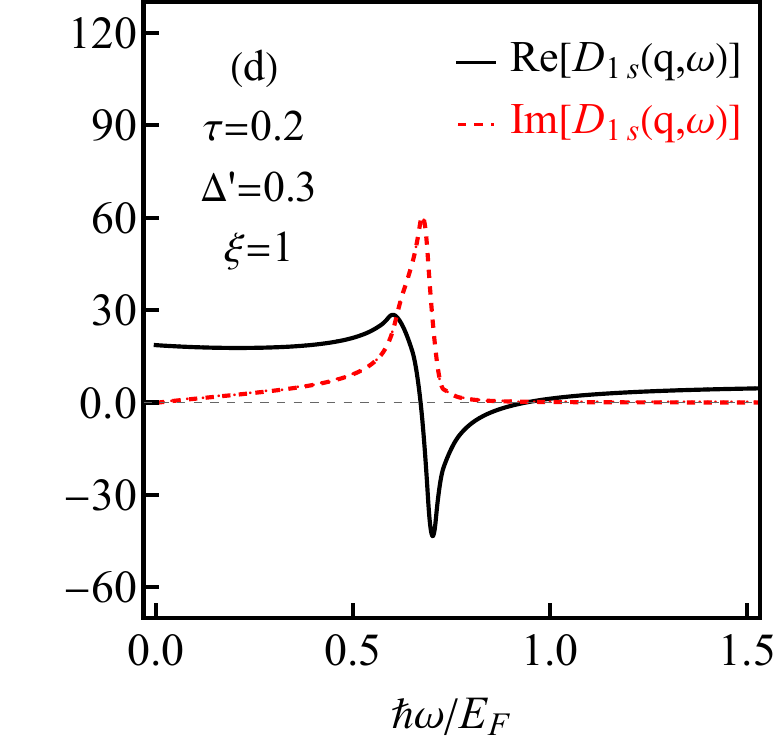}
\label{d_w4}
\end{subfigure}
\captionsetup{singlelinecheck=off,justification=raggedright}
\caption{(Color online) Real and imaginary parts of  $D_{1s}({\bf q},\omega|E_{F})$ for a monolayer placed on a dielectric  substrate when $q_{x}=0$ and $q_{y}=0.6$. The plots are plotted with respect to frequency $\omega$ in unit of $E_{F}/\hbar$. The dielectric constant of the substrate is $\epsilon_{1}=5.0$ and the dephasing parameter chosen is $\hbar\delta^{+}/E_{F}=0.01$ as that in Fig.~\ref{expansion} and Fig.~\ref{density1qxqy_re}. The black curve depicts the real part, while the red dashed line represents the imaginary part of the dielectric function. Panel (a) is the dielectric function when there is no band gap or tilting, (b) is the dielectric function when gap $\Delta/E_{F}=0.3$ is present, (c) is the dielectric function when tilting $\tau=0.2$ is present and (d) is the dielectric function when both a band gap $\Delta/E_{F}=0.3$ and tilting $\tau=0.2$ are included.}
\label{dielectric}
\end{figure}
\medskip
\par

Figure.~\ref{dielectric} presents the real and imaginary parts of the dielectric function for a monolayer of gapped TSDM placed on a thick dielectric substrate as a function of frequency $\omega$ which is scaled in unit of $E_{F}/\hbar$. These results are also based on the exact  expression of $D_{1s}({\bf q},\omega|E_{F})$ given by Eq.~(\ref{D1}), which is proportional to the dielectric function for one layer placed on a substrate, and is not on the long wavelength approximation. The small dephasing parameter added to the frequency is $\hbar\delta^{+}/E_{F}=0.01$ and the dielectric constant of the substrate is $\epsilon_{1}=5.0$, same as that in Fig.~\ref{density1qxqy_re}. The chosen values for $q_{x}/k_{F}$ and $q_y/k_{F}$ are $0.0$ and $0.6$, respectively. Figure.~\ref{d_w1} is plotted when there is no tilting or energy band gap. The black curve and dashed red line depict the real and imaginary parts of the dielectric function, respectively. We observe that when the real part first crosses the frequency axis, it does so in the particle-hole mode region, at that frequency, the imaginary part has a peak. This indicates the presence of strong Landau damping of the plasmon mode by single-particle excitations. We notice that the real part of the dielectric function crosses the frequency axis a second time at an undamped plasmon frequency. Figure.~\ref{d_w2} presents the dielectric function when the gap $\Delta '=0.3$ is chosen, and we notice that the height of the peak for the imaginary part is decreased, thereby indicating that the Landau damping of the plasmon mode is reduced. When there is tilting $\tau=0.2$, Fig.~\ref{d_w3} shows that not only the height of the peak is slightly reduced, but the plasmon frequency is increased. Figure~\ref{d_w4} clearly shows the reduction in heights of the peaks and increase in plasmon frequency when both the band gap and tilting are present. Therefore, in the presence of energy gap, $\Delta '$=0.3, the damping is reduced, but we still observe large landau damping that causes the first cross of the frequency axis of the real part to be particle-hole mode. The tilting $\tau$=0.2 causes the plasmon frequency to increase, however, imaginary part of the dielectric function is still zero when the real part of the plasmon frequency crosses second time. Therefore, this Fig.~\ref{dielectric} verifies the Fig.~\ref{density1qxqy_re} which shows only one plasmon branch, when tilting $\tau$=0.1 and gap $\Delta '$=0.1 are present with small dephasing parameter of $\hbar\delta^{+}/E_{F}=0.01$.


\subsection{Comparison between plasmons for two layers embedded in a dielectric and coating a dielectric on a substrate}

\begin{figure}[h]
\centering
\begin{subfigure}{.45\textwidth}
\refstepcounter{subfigure}
\centering
\includegraphics[width=.95\columnwidth]{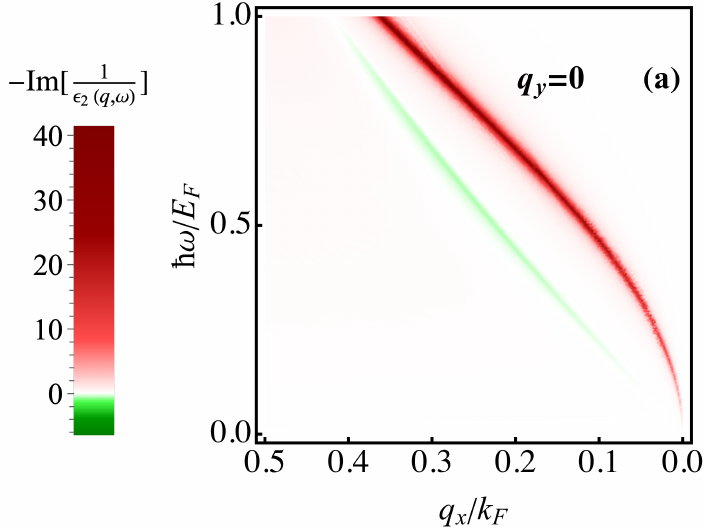}
\caption*{}
\label{dqx} 
\end{subfigure}%
\begin{subfigure}{.45\textwidth}
\refstepcounter{subfigure}
\centering
\includegraphics[width=.95\columnwidth]{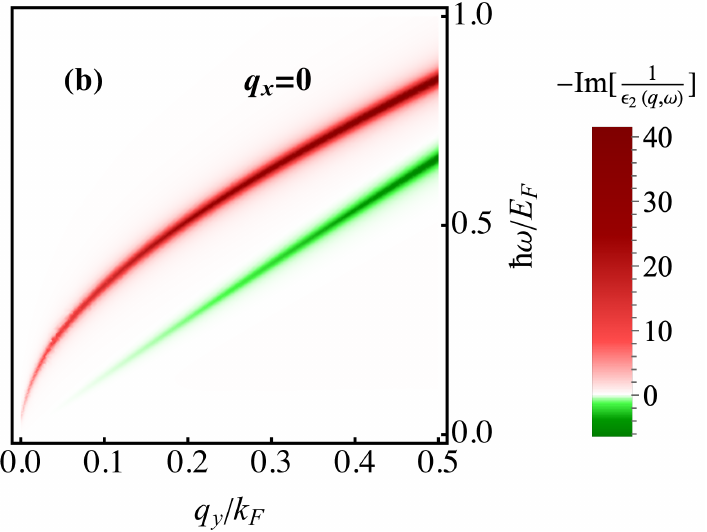}
\caption*{}
\label{dqy}
\end{subfigure}
\captionsetup{singlelinecheck=off,justification=raggedright}\caption{(Color online) Density plots for the imaginary parts of $-1/\epsilon_{2}({\bf q},\omega)$ for a pair of gapped TSDM monolayers embedded in a uniform background dielectric with $\epsilon_{1}=4.0$. We chose tilting $\tau=0.1$, energy gap parameter $\Delta'=0.1$, and dephasing parameter as $\hbar\delta^+/E_F=0.01$. The separation between the pair of monolayers is $d_{1}/d_{0}=1.5$ where it is in unit of $d_{0}=1/k_{F} \approx 10^{-8}$m. The penal (a) is the dispersion frequency $\omega$ in unit of $E_{F}/\hbar$, plotted as a function of $q_{x}/k_{F}$ when $q_{y} =0$, while (b) is the result plotted with respect to $q_y/k_F$ when $q_x =0$. The red color indicates a weakly damped mode, and the green color indicates a strongly damped mode. Also the transition of color from white to dark red represents the reduction in damping and the transition of color from white to dark green represents the increase in damping.}
\label{determinant}
\end{figure}

We further investigate the difference between plasmonic properties of this standard double-layered geometry and the coated structure placed on a substrate by plotting the plasmon dispersions. In Fig.~\ref{determinant}, we first present density plots for the loss function of the standard double layered structure, where the dielectric function, $\epsilon_{2}({\bf q},\omega)$, is given by Eq.~(\ref{sarma}). We also present density plots for the imaginary part of $-\frac{1}{D_{2s}}$ in Fig.~\ref{substrate}, where $D_{2s}$ is given by Eq.~(\ref{D2s}), which is the denominator of the SRF for a coated structure placed on a substrate.

\begin{figure}[h]
\centering
\begin{subfigure}{.45\textwidth}
\refstepcounter{subfigure}
\centering
\includegraphics[width=.95\columnwidth]{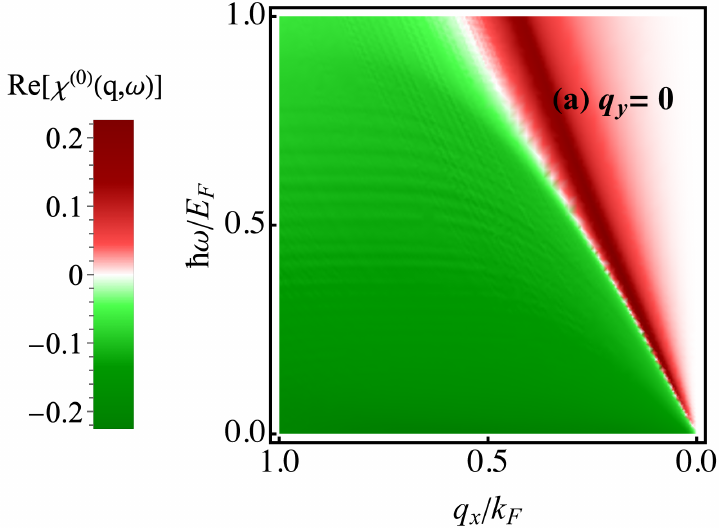}
\caption*{}
\label{pol_qx1} 
\end{subfigure}%
\begin{subfigure}{.45\textwidth}
\refstepcounter{subfigure}
\centering
\includegraphics[width=.95\columnwidth]{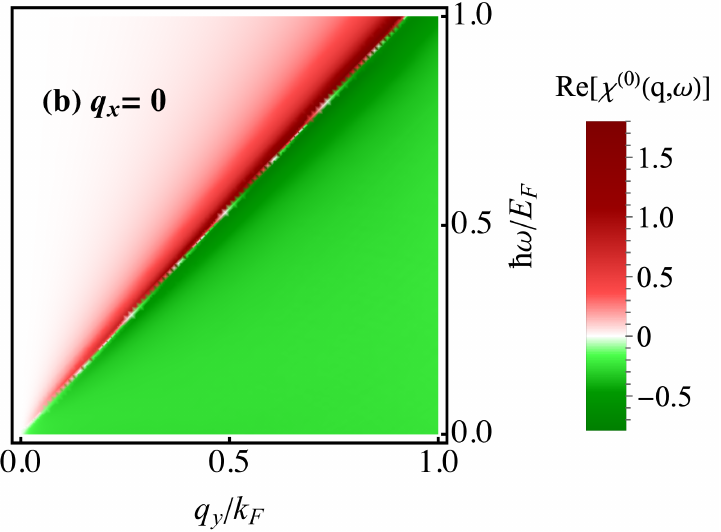}
\caption*{}
\label{pol_qy1}
\end{subfigure}
\begin{subfigure}{.45\textwidth}
\refstepcounter{subfigure}
\centering
\includegraphics[width=.95\columnwidth]{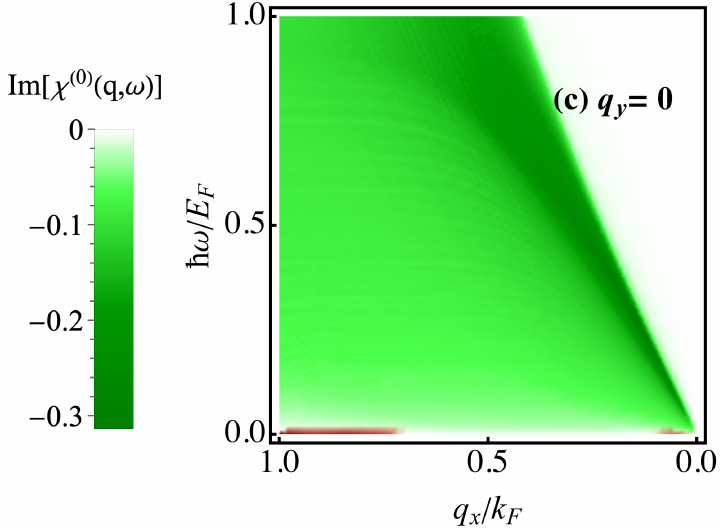}
\caption*{}
\label{pol_qx2} 
\end{subfigure}%
\begin{subfigure}{.45\textwidth}
\refstepcounter{subfigure}
\centering
\includegraphics[width=.95\columnwidth]{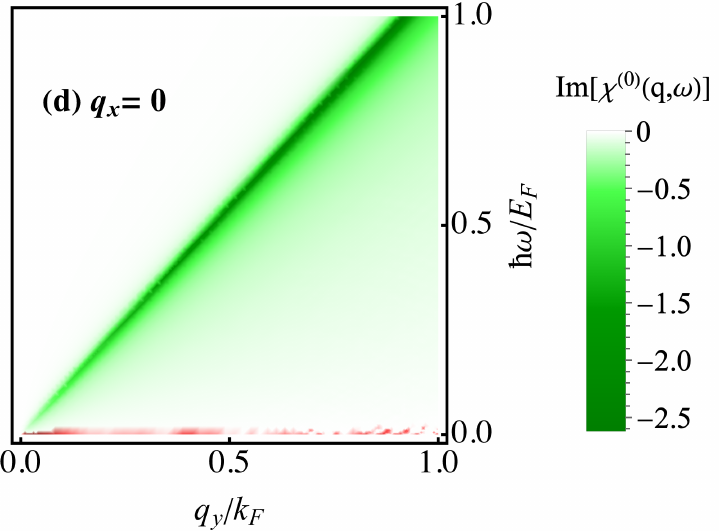}
\caption*{}
\label{pol_qy2}
\end{subfigure}
\captionsetup{singlelinecheck=off,justification=raggedright}\caption{(Color online) (a) and (b) are density plots for the real parts of the polarization function $\chi^{(0)}(\bf{q,}\omega)$ for gapped TSDM where tilting parameter $\tau=0.1$, energy gap parameter $\Delta'=0.1$, and dephasing parameter are $\hbar\delta^+/E_F=0.01$. Plots (c) and (d) are the imaginary part of the polarization function $\chi^{(0)}(\bf{q,}\omega)$. Plots (a) and (c) shows $\omega$ in unit of $E_{F}/\hbar$ as functions of $q_{x}/k_{F}$ when $q_{y} =0$, and (b) and (d) are plotted with respect to $q_y/k_F$ when $q_x =0$. The red color represents positive values and the green color represents negative values of the real or imaginary parts of the polarization function.}
\label{pol}
\end{figure}

\begin{figure}[h]
\centering
\begin{subfigure}{.45\textwidth}
\refstepcounter{subfigure}
\centering
\includegraphics[width=.95\columnwidth]{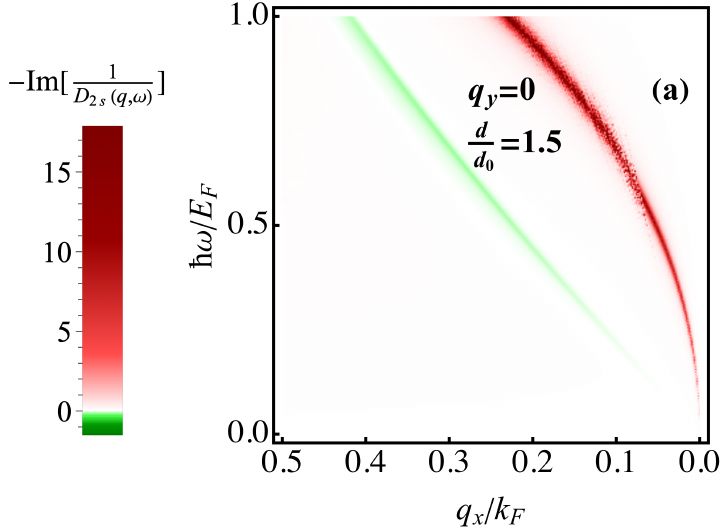}
\caption*{}
\label{substrate_qx1} 
\end{subfigure}%
\begin{subfigure}{.45\textwidth}
\refstepcounter{subfigure}
\centering
\includegraphics[width=.95\columnwidth]{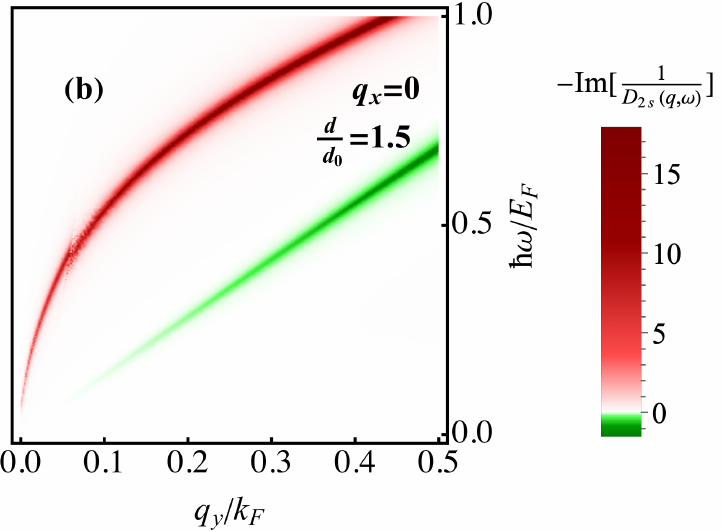}
\caption*{}
\label{substrate_qy1}
\end{subfigure}
\begin{subfigure}{.45\textwidth}
\refstepcounter{subfigure}
\centering
\includegraphics[width=.95\columnwidth]{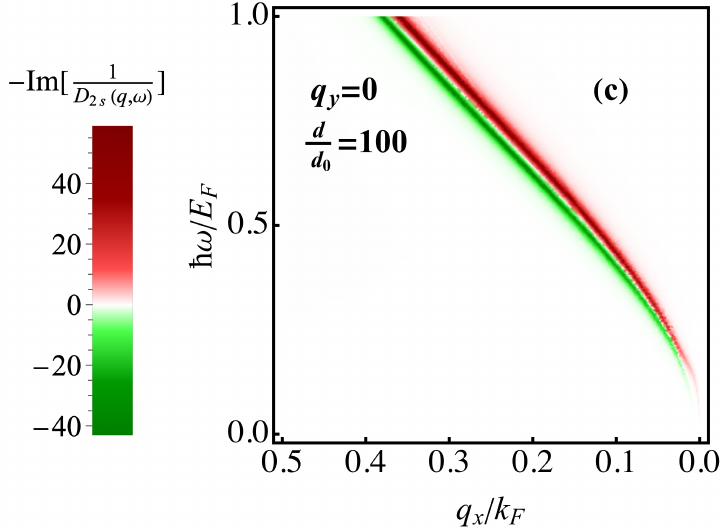}
\caption*{}
\label{substrate_qx2} 
\end{subfigure}%
\begin{subfigure}{.45\textwidth}
\refstepcounter{subfigure}
\centering
\includegraphics[width=.95\columnwidth]{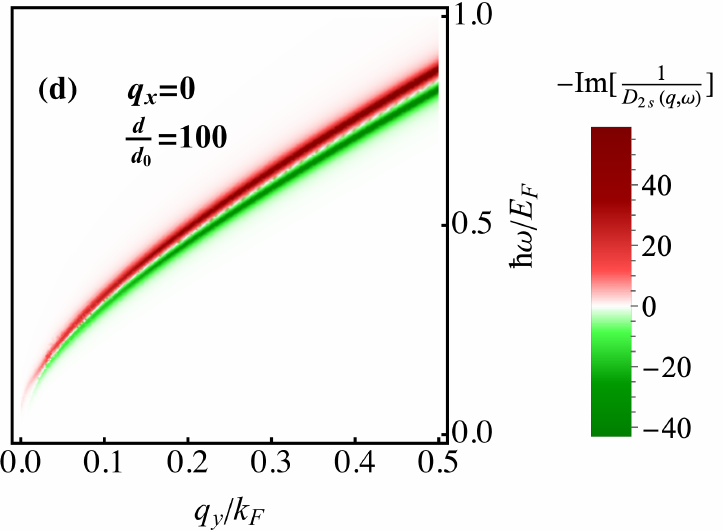}
\caption*{}
\label{substrate_qy2}
\end{subfigure}
\captionsetup{singlelinecheck=off,justification=raggedright}\caption{(Color online) Density plots for the imaginary parts of $-1/D_{2s}({\bf q},\omega)$ for a pair of gapped TSDM monolayers encapsulating a dielectric material with $\epsilon_{b}=4.0$ and placed on a substrate with dielectric constant $\epsilon_{2}=2.0$. We chose tilting parameter $\tau=0.1$, energy gap parameter $\Delta'=0.1$, and dephasing parameter $\hbar\delta^+/E_F=0.01$. The plots (a) and (b) have separation between the pair of monolayers $d_{1}/d_{0}=1.5$ where it is in unit of $d_{0}=1/k_{F} \approx 10^{-8}$m.  Plot (a) is the dispersion frequency $\omega$ in unit of $E_{F}/\hbar$ as a function of $q_{x}/k_{F}$ when $q_{y} =0$, and (b) shows $\omega$ in unit of $E_F/\hbar$ plotted with respect to $q_y/k_F$ when $q_x =0$. (c) and (d) are the plots for $d_{1}/d_{0}=100$ where (c) is the plot for $\omega$ as a function of $q_{x}/k_{F}$ when $q_{y} =0$, and (d) is plot for $\omega$ as a function of $q_y/k_F$ when $q_x =0$.}
\label{substrate}
\end{figure}

\medskip
\par
In the plasmon dispersion in Fig.~\ref{determinant}, the two layers of gapped TSDM with tilting of magnitude $\tau=0.1$ and energy gap parameter of magnitude $\Delta'=0.1$ and dephasing parameter of $\hbar\delta^{+}/E_{F}=0.01$ are embedded in a uniform dielectric medium with an intermediate background value of $\epsilon_{b}=4.0$, which lies between the reported in-plane (6.82 to 6.93) and out-of plane (3.29 to 3.76) dielectric constants of h-BN\cite{Akash}. The plasmon dispersion has an anisotropic dependence on the wave vector components $q_x$ and $q_y$ such that the excitation energy is slightly higher in the $q_{x}$ direction compared with the $q_{y}$ direction. Of course, this is a reflection of the anisotropy and tilting of the energy bands for TSDM. Moreover, it is worthy of note that we have weakly damped optical branches and highly damped acoustic branches indicated by red and green colors respectively, referring to the side-bar code in Fig.~\ref{determinant}. To clarify this behavior, we plotted Fig.~\ref{pol}, the real and imaginary parts of the polarization function for gapped TSDM with energy gap parameter of magnitude $\Delta'=0.1$, tilting of magnitude $\tau=0.1$, and dephasing parameter of $\hbar\delta^{+}/E_{F}=0.01$. In this case, red color represents positive values and green color represents negative values of real or imaginary parts of the polarization function. We notice that the imaginary part of the polarization function is always on the negative side indicated by green color in both $q_{x}$ and $q_{y}$ directions as shown in Fig.~\ref{pol_qx2} and Fig.~\ref{pol_qy2}. Therefore, we are aware that the density plot of the loss function for monolayer would correspond to Landau damping as shown in Fig.~\ref{density1qxqy_re} since the loss function is the negative of the imaginary part of the inverse dielectric function. However, for two layers, the dispersion function involves the product of two polarization functions, one from each layer as easily verified from Eq.~(\ref{sarma}). While the imaginary part of the polarization function is always on the negative side, the sign  of the real part of the polarization function can vary with wave vector as shown in Fig.~\ref{pol_qx1} and Fig.~\ref{pol_qy1}, since the real part can cross the frequency axis for fixed wave vector which is the behavior required for the determination of the plasmon dispersion. Therefore, the loss function for two layers involving the product of the imaginary part and real part of the polarization function would yield a side bar code with weakly Landau damped and highly damped modes, here  indicated by red and green color, respectively. In the color code bar, the transition of color from white to dark red indicates the reduction in damping. Conversely, the transition of color from white to dark green indicates strong damping. As shown in Fig.~\ref{determinant}, weakly damped high-frequency optical branches and the highly damped lower-frequency acoustic branches convey the well known characteristic of the optical mode which is in in-phase, and the acoustic mode is out-of-phase\cite{Hwang}. We farther note that the strength of the damping is nonuniform across the spectrum. For instance, in Fig.~\ref{dqy}, the dampings of the optical branches become weaker as the wave vectors increase, as we can see the transition of color from white to dark red. Conversely, the dampings of acoustic branches become stronger as the wave vectors increase, as we can observe the transition of color from white to darker green. We also notice that the magnitudes of imaginary or real parts of the polarization functions vary along the chosen wave vectors, especially in Fig.~\ref{pol_qy2}, we can observe that imaginary part of the polarization function changes along the linear boundary as the wave vector $q_{y}$ increases. 

\medskip
\par

In summary, our results are consistent with those reported in Ref.\cite{Sarma} for two 2D layers embedded in a uniform dielectric background. Besides this, we have explicitly demonstrated that the existing two plasmon branches can have different levels of damping which may play a role in experimental detection and potential applications. In this case, we plot the weakly damped optical branch with red color and highly damped acoustic branch with green color. 
\medskip
\par

\medskip
\par
Let us now turn our attention to the plots in Fig.~\ref{substrate} which, unlike Fig.~\ref{determinant}, are plasmon dispersions for coated structures with vacuum on one side and placed on a substrate with dielectric constant $\epsilon_{2}=2.0$. The average dielectric constant values in the range of 2.5 - 3.1 are observed for poly  (methyl methacrylate) dissolved in cholorobenzene or anisole, depending on the solvent concentration \cite{Estrada}. Figure.~\ref{substrate_qx1} and Fig.~\ref{substrate_qy1} refer to layers having the same separation as that chosen in Fig.~\ref{dqx} and Fig.~\ref{dqy}. Here, the same TSDM monolayers are used in the calculations and the monolayers are employed for coating the same dielectric material as in Fig.~\ref{determinant}. The anisotropic behavior of the plasmon branches in Figs.~\ref{substrate_qx1} through Fig.~\ref{substrate_qy2} is very clear. The plasmon dispersion in the $q_{y}$ direction is extended slightly farther than in $q_{x}$ direction. Additionally, the separation between the optical and acoustic branches in Fig.~\ref{substrate_qx1} and Fig.~\ref{substrate_qy1} is significantly larger than in Fig.~\ref{dqx} and Fig.~\ref{dqy}, while the lifetimes for weakly and highly damped modes are both suppressed, thereby demonstrating that the surrounding media have a nontrivial influence on the screening that influences the plasmon dispersion. We would also like to highlight that our calculations show that when the separation between layers is $d/d_{0}=1.5$, as shown in Figs.~\ref{dqx}, \ref{dqy}, \ref{substrate_qx1} and \ref{substrate_qy1}, the acoustic and optical branches are clearly separated. However when the distance between the monolayers is increased from $d/d_{0}=1.5$ to $d/d_{0}=100.0$, we observe that the optical and acoustic branches are now packed together, and the lifetimes of weakly and highly damped modes are modified as depicted in Fig.~\ref{substrate_qx2} and Fig.~\ref{substrate_qy2}. Also, we observe that spectral lifetimes of two modes are now comparable. This indicates that, after increasing the separation between the monolayers, the optical branches became more weakly damped and the acoustic branches became more highly damped at a similar rate, strengthening the characteristics of the in-phase mode for the optical branch and out-of phase mode for the acoustic branch. Das Sarma and Madhukar [\onlinecite{Sarma}] state that when the separation between the two layers of charge carriers in bulk is increased, the interaction between the two components is reduced, which affect the acoustic branch. Therefore, our plots clearly verify the validity of their assertion. Moreover, compared with the standard  bulk model with two monolayers embedded in a uniform background material, the plasmon excitation frequencies of the branches in the coated structure are higher, indicating that they require more energy to excite. Therefore, we have discovered that the arrangement with vacuum above and a substate below the coated sample increases the energy required to excite plasmons than when the two monolayers are embedded in a bulk material. 
\medskip
\par

Now we take the long wavelength limit of the plasmon dispersion relation for the coated structure resting on a thick substrate. Setting $D_{2s}(\mathbf{q},\omega)=0$ in Eq.~(\ref{D2s}), we obtain two plasmon branches in two mutually perpendicular directions  along $q_x$ and $q_y$ as 
\begin{equation}
\bigr(\hbar\omega_{p}^{\pm}\bigr)^{2}\approx \frac{1}{2S_{0}(\mathbf{q})}\biggr(-S_1(\mathbf{q}) \pm \sqrt{S_{1}^2(\mathbf{q}) -4 S_0(\mathbf{q}) 
S_2(\mathbf{q})}\biggr)
\label{2modes}
\end{equation}
with
\begin{equation}
\begin{split}
&S_0(\mathbf{q})= e^{-2qd_{1}}(-\epsilon_{1}\epsilon_{2}+\epsilon_{2}+\epsilon_{1}^{2}-\epsilon_{1})+(-\epsilon_{1}\epsilon_{2}-\epsilon_{2}-\epsilon_{1}^{2}-\epsilon_{1})\\
&S_1(\mathbf{q})=\frac{e^{2}}{\epsilon_{0}q}e^{-2qd_{1}}\Biggr[-\epsilon_{2}P^{(0)}({\mathbf{q}})+\epsilon_{1}P^{(1)}({\mathbf{q}})-P^{(1)}({\mathbf{q}})+\epsilon_{1}P^{(0)}({\mathbf{q}})\Biggr] \\
&+\frac{e^{2}}{\epsilon_{0}q}\Biggr[\epsilon_{2}P^{(0)}({\mathbf{q}})+\epsilon_{1}P^{(1)}({\mathbf{q}})+P^{(1)}({\mathbf{q}})+\epsilon_{1}P^{(0)}({\mathbf{q}})\Biggr]  \\
&S_2(\mathbf{q})= \biggr(\frac{e^2}{\epsilon_0q} \biggr)^{2} \Bigr(P^{(0)}(\mathbf{q}) \cdot P^{(1)}({\mathbf{q}})\Bigr)(e^{-2qd_{1}}-1)
\label{2modes2}
\end{split}
\end{equation}
The expressions for polarizations in long wavelength limit, $P^{(0)}$ and $P^{(1)}$ are given in Eq.~(\ref{Aeqn}). Both $P^{(0)}$ and $P^{(1)}$  determine the polarization functions of the upper layer and lower layer TSDMs respectively. In principle, $P^{(1)}$ can have the same or different energy gap parameter $\Delta/E_{f}$ from $P^{(0)}$. 
\begin{figure}[h]
\centering
\begin{subfigure}{.38\textwidth}
\refstepcounter{subfigure}
\includegraphics[width=.85\columnwidth]{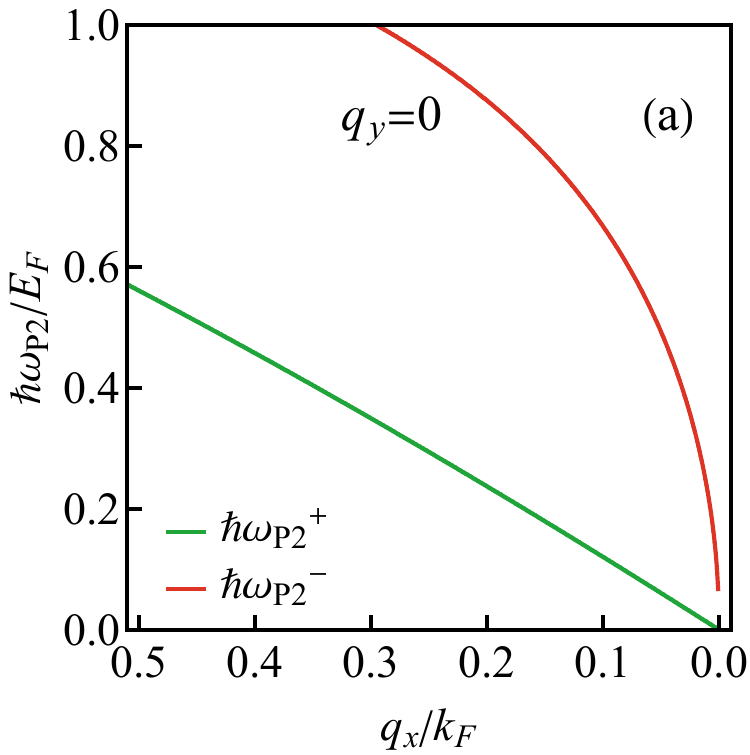}
\caption*{}
\label{d2qx} 
\end{subfigure}%
\hspace{0pt}
\begin{subfigure}{.38\textwidth}
\refstepcounter{subfigure}
\includegraphics[width=.85\columnwidth]{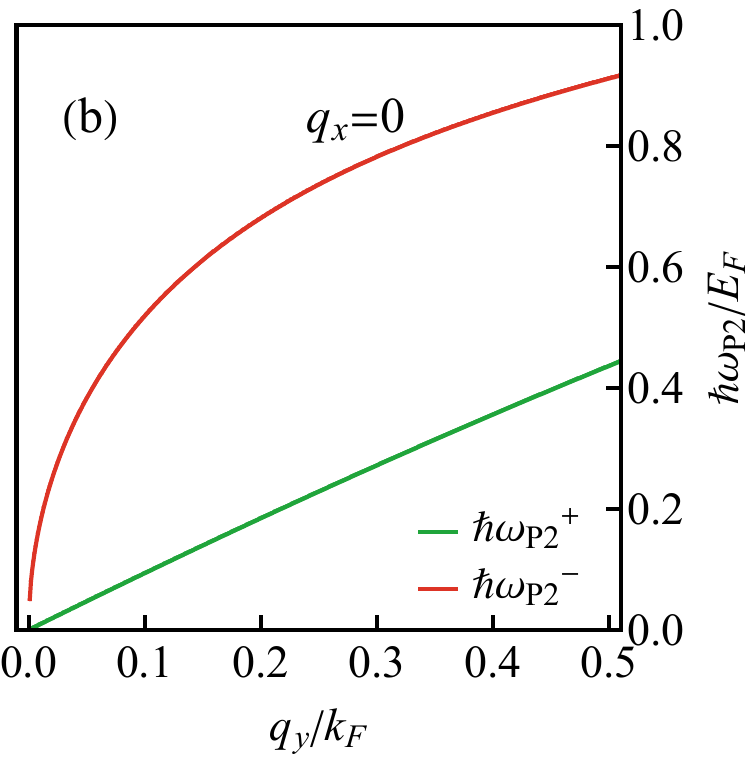}
\caption*{}
\label{d2qy}
\end{subfigure}\captionsetup{singlelinecheck=off,justification=raggedright}
\caption{(Color online) (a) Plasmon dispersion in the  long wavelength limit for two monolayers surrounding a dielectric $\epsilon_{1}=4.0$, and placed on a thick substrate with dielectric constant $\epsilon_{2}=2.0$. Both top and bottom electron layers have tilting $\tau=0.1$ and energy gap parameter $\Delta'=0.1$.  The separation between  the two monolayers is $d/d_{0}=$ 1.5, in unit of $d_{0} = \frac{1}{k_F}$. The red line shows the higher frequency branch while the  green line  is for lower frequency modes. In (a),  the frequency is in unit of $E_{F}/\hbar$ and is plotted as a function  of $q_{x}/k_{F}$ when $q_{y}  =0$. In (b),  the plasmon frequencies are plotted as  a function of $q_{y}/k_{F}$ when $q_{x} =0$. }
\label{d2}
\end{figure}
The Fig.~\ref{d2} displays the plot of plasmon frequency in the long wavelength limit, when two layers have the same energy gap parameter of $\Delta'=0.1$, tilting $\tau=0.1$, and separated by a distance of $d_{1}/d_{0}=1.5$ where it is in unit of $d_{0}=1/k_{F}\approx 10^{-8}$m. The frequency is in unit of $E_{F}/\hbar$ where $E_{F}=\hbar\upsilon_{F}k_{F}$ and the Fermi velocity $\upsilon_{F}=10^6$m/s is used. The Fig.~\ref{d2qx} is plotted with respect to $q_x/k_{F}$ when $q_y=0$. Figure.~\ref{d2qy} is plotted with respect to $q_y/k_{F}$ when $q_x =0$ and we note that  two plasmon branches appear. Comparing the plots in two perpendicular directions, we also observe that the plasmon branches in the $q_x $ direction are slightly higher than those in the $q_y$ direction. This demonstrates the existence of anisotropy.

\subsection{Comparison of  homogeneous and inhomogeneous dielectric background on plasmons for three gapped TSDM monolayers}

\begin{figure}[h]
\centering
\begin{subfigure}{.45\textwidth}
\refstepcounter{subfigure}
\centering
\includegraphics[width=.95\columnwidth]{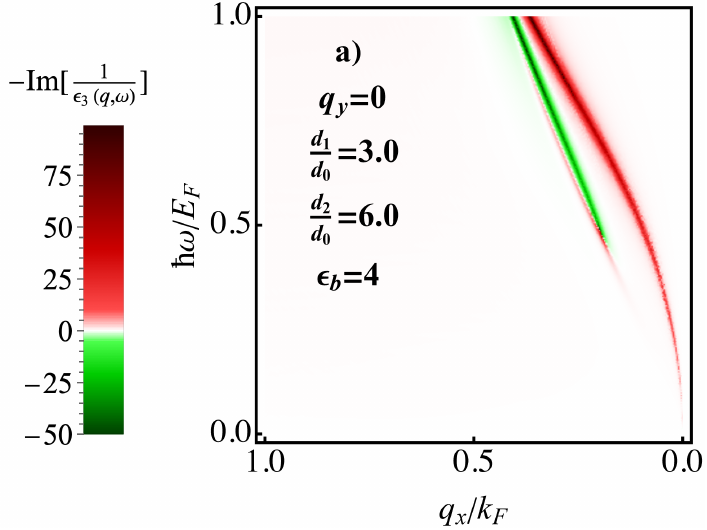}
\caption*{}
\label{threeb_qx1} 
\end{subfigure}%
\begin{subfigure}{.45\textwidth}
\refstepcounter{subfigure}
\centering
\includegraphics[width=.95\columnwidth]{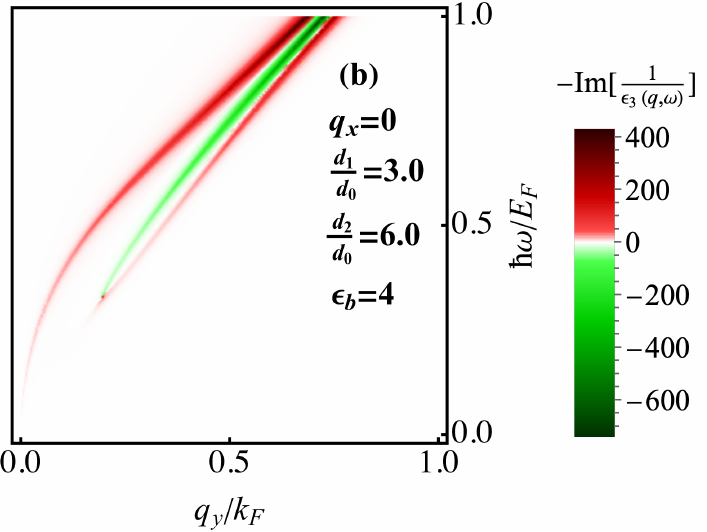}
\caption*{}
\label{threeb_qy1}
\end{subfigure}
\begin{subfigure}{.45\textwidth}
\refstepcounter{subfigure}
\centering
\includegraphics[width=.95\columnwidth]{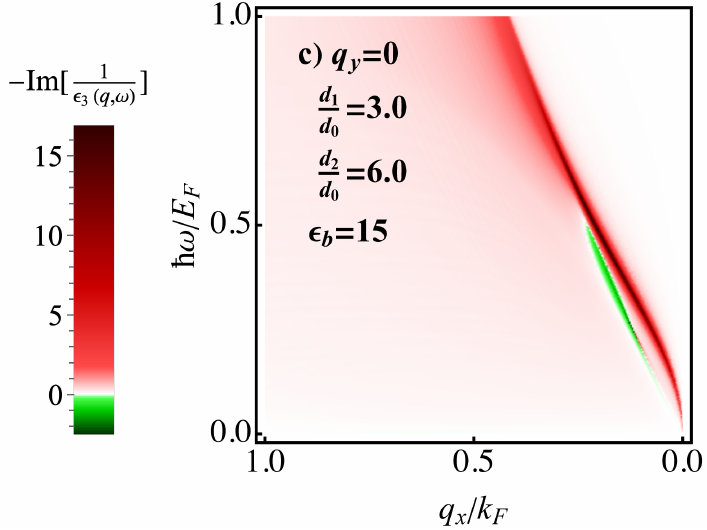}
\caption*{}
\label{threeb_qx2} 
\end{subfigure}%
\begin{subfigure}{.45\textwidth}
\refstepcounter{subfigure}
\centering
\includegraphics[width=.95\columnwidth]{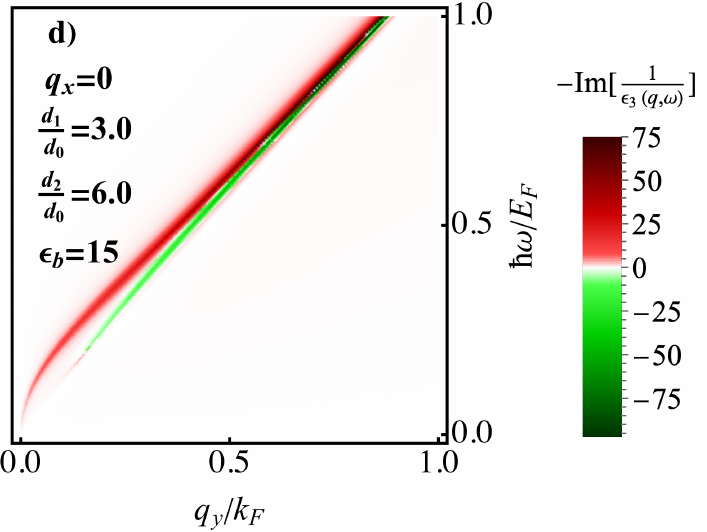}
\caption*{}
\label{threeb_qy2}
\end{subfigure}
\captionsetup{singlelinecheck=off,justification=raggedright}\caption{(Color online) Density plots for the imaginary parts of $-1/\epsilon_{3}({\bf q}\omega)$ for three layers of gapped TSDM monolayers with tilting $\tau=0.1$, band gap parameter $\Delta'=0.1$, and dephasing parameter $\hbar\delta^+/E_F=0.01$ embedded in a homogeneous dielectric material. The structure have separations between the first pair of monolayers $d_{1}/d_{0}=3.0$, and the third layer from the first layer $d_{2}/d_{0}=6.0$ which in unit of $d_{0}=1/k_{F} \approx 10^{-8}$m. (a) and (b) are with homogeneous dielectric material $\epsilon_{b}=4.0$ and plots (c) and (d) are that with increased $\epsilon_{b}=15$. Plots (a), (c) show the dispersion frequency $\omega$ in unit of $E_{F}/\hbar$ as a function of $q_{x}/k_{F}$ when $q_{y} =0$, and (b), (d) show $\omega$ in unit of $E_F/\hbar$ plotted with respect to $q_y/k_F$ when $q_x =0$. } 
\label{threelayers_eb}
\end{figure}

\medskip
\par
We now discuss and contrast the plasmonic properties for two cases when three monolayers of gapped TSDMs are embedded in a homogeneous dielectric medium and when these three monolayers are separated  by two different dielectric materials and this system is then suspended in air. For both cases, the dephasing factor is chosen as $\hbar \delta^{+}/E_{F}=0.01$. The expression in Eq.~(\ref{3vacuum}) is used to plot the plasmonic dispersion of three tilted and gapped TSDM monolayers embedded in a background dielectric with magnitude of $\epsilon_{b}=4$ as shown in Fig.~\ref{threeb_qx1} and Fig.~\ref{threeb_qy1}. Here, three gapped TSDM monolayers have the same tilting parameter $\tau=0.1$ and energy gap parameter $\Delta'=0.1$. The separation between the first and second layers is $d_{1}/d_{0}=3.0$, and the separation between the first and third layers is $d_{2}/d_{0}=6.0$. In both $q_{x}$ and $q_{y}$ directions, the optical-like highest branches are on the weakly damped side and clearly observed while the second branches are on the highly damped side of the color code, and third branches are on the weakly damped side but with very small reduction of damping such that it's hard to observe especially in $q_{x}$ direction. In Fig.~\ref{threeb_qx1} and \ref{threeb_qy1}, we can observe two lower branches merged together at critical $q_{x}^{c}$ and $q_{y}^{c}$ in both $q_{x}$ and $q_{y}$ directions. Two branches in each direction then separate/anticross as the corresponding wave vectors are increased. As the magnitude of the background dielectric medium is increased from $\epsilon_{b}=4.0$ to $\epsilon_{b}=15.0$ as shown in Fig.~\ref{threeb_qx2} and Fig.~\ref{threeb_qy2}, we observe that the critical $q_{c}$ is now much lower and the bottom most branches are almost hard to observe.

\begin{figure}[h]
\centering
\begin{subfigure}{.45\textwidth}
\refstepcounter{subfigure}
\centering
\includegraphics[width=.95\columnwidth]{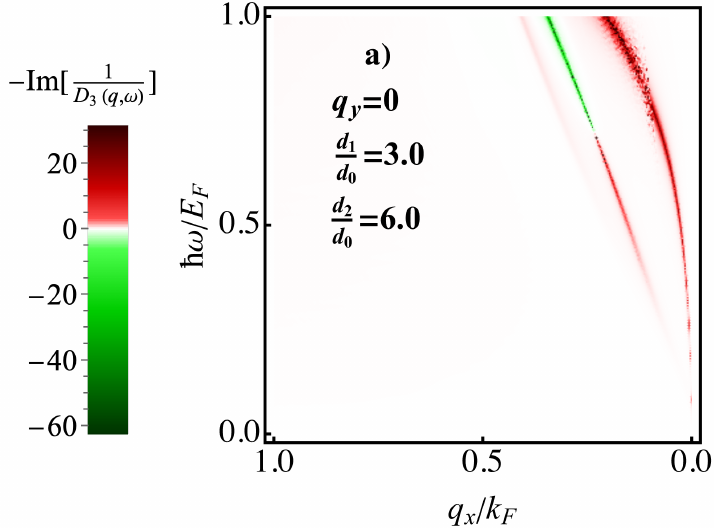}
\caption*{}
\label{three_qx1} 
\end{subfigure}%
\begin{subfigure}{.45\textwidth}
\refstepcounter{subfigure}
\centering
\includegraphics[width=.95\columnwidth]{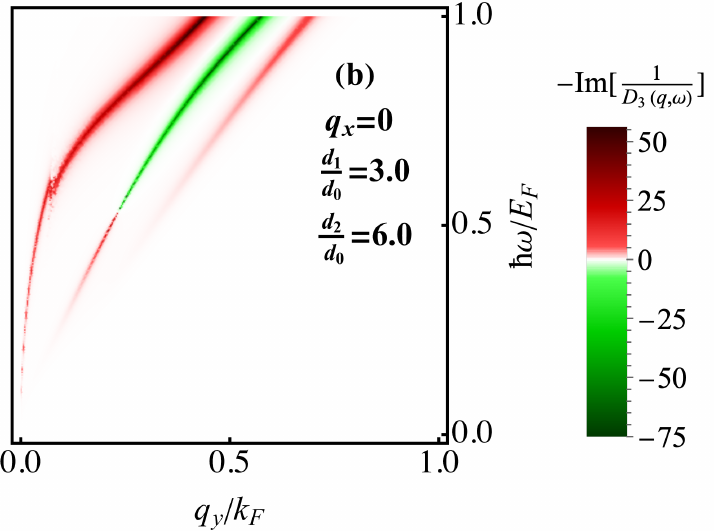}
\caption*{}
\label{three_qy1}
\end{subfigure}
\begin{subfigure}{.45\textwidth}
\refstepcounter{subfigure}
\centering
\includegraphics[width=.95\columnwidth]{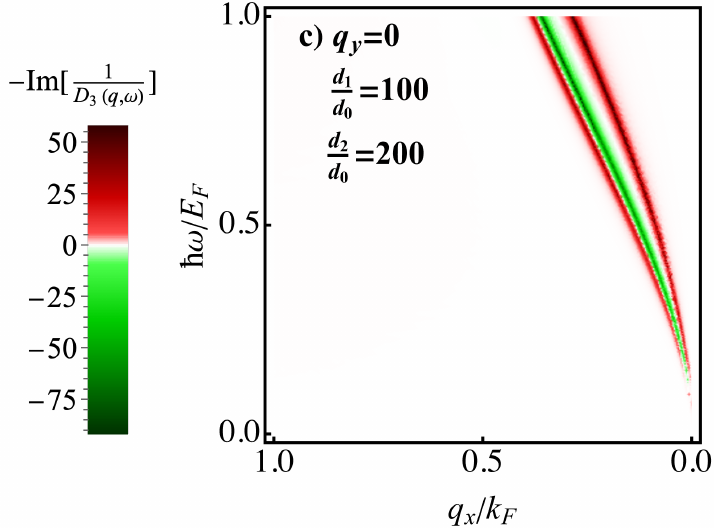}
\caption*{}
\label{three_qx2} 
\end{subfigure}%
\begin{subfigure}{.45\textwidth}
\refstepcounter{subfigure}
\centering
\includegraphics[width=.95\columnwidth]{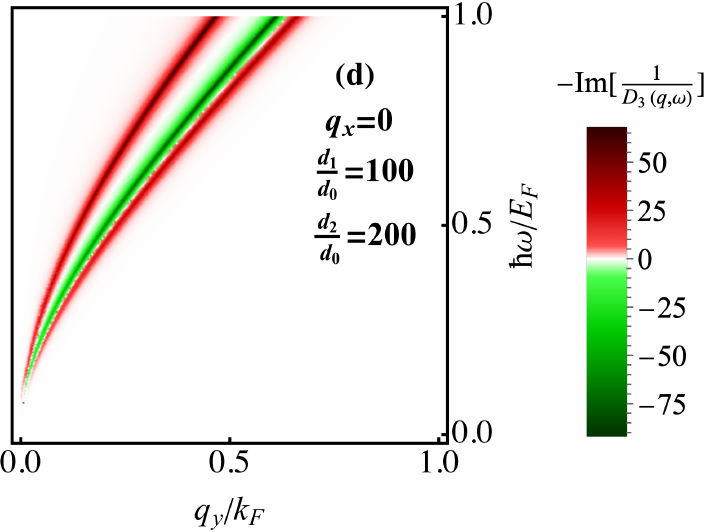}
\caption*{}
\label{three_qy2}
\end{subfigure}
\captionsetup{singlelinecheck=off,justification=raggedright}\caption{(Color online) Density plots for the imaginary parts of $-1/D_{3}({\bf q},\omega)$ for three layers of gapped TSDM monolayers encapsulating a first dielectric material with $\epsilon_{1}=4.0$ and second dielectric material with $\epsilon_{2}=2.0$. We chose tilting $\tau=0.1$, The band gap parameter is $\Delta^\prime=0.1$ and the dephasing parameter $\hbar\delta^+/E_F=0.01$. The plasmon branches in (a) and (b) have separations between the first pair of monolayers $d_{1}/d_{0}=3.0$, and the third layer from the first layer $d_{2}/d_{0}=6.0$  in unit of $d_{0}=1/k_{F} \approx 10^{-8}$m. Plot (a) is the dispersion frequency $\omega$ in unit of $E_{F}/\hbar$ is as a function of $q_{x}/k_{F}$ when $q_{y} =0$, and (b) shows $\omega$ in unit of $E_F/\hbar$ plotted with respect to $q_y/k_F$ when $q_x =0$. (c) and (d) are plots for $d_{1}/d_{0}=100$ and  $d_{2}/d_{0}=200.0$ whereas (c) is the plot for $\omega$ as a function of $q_{x}/k_{F}$ when $q_{y} =0$, and (d) is plots for $\omega$ as a function of $q_y/k_F$ when $q_x =0$.}
\label{threelayers}
\end{figure}

\medskip
\par

\medskip
\par
We are now in a position to compare from a qualitative point of view the results for the plasmons in a three-layer 2D structure of gapped TSDMs with a homogeneous versus inhomogeneous dielectric background. The differences between Fig.~\ref{threelayers_eb} and Fig.~\ref{threelayers} are stark. We emphasize that there is no tunneling between layers or hybridization of the polarization function. As shown in Figs.~\ref{three_qx1} and \ref{three_qy1}, for the case of inhomogeneous dielectric background, when the separation between the first and second layer is $d_{1}/d_{0}=3.0$ and the first and third layer is $d_{2}/d_{0}=6.0$, we can clearly detect that the highest branches are very weakly damped in both directions. The damping of the second highest branches vary with the change in wave vectors $q_x$ and $q_y$ in a non-uniform way. First, when $q_x$ and $q_y$ are small, the second highest branches begin as weakly damped modes with light red color, then the damping become weaker as the wave vectors increase, however, at certain points, they start to gain damping, such that they change to highly damped mode as indicated by dark green color. The third lowest branches, which are in the weakly damped mode, the dampings become weaker and weaker as the $q_x$ and $q_y$ increase. The overall change in the extent of damping is due to the product of imaginary and real parts of the polarization functions which are involved in the expression for the SRF. Now, when we increased the separation between the first and second layer to $d_{1}/d_{0}=100.0$, as well as the first and third layer to $d_{2}/d_{0}=200.0$, we can observe three branches more clearly as shown in Figs. ~\ref{three_qx2} and \ref{three_qy2}. The highest branches are weakly damped, the second highest branches are highly damped and the lowest branches are weakly damped and their visibility is comparable. As we increased the separations between the layers, we notice that the distance between the three branches becomes more packed together and the overall structure becomes more linear. Overall, compared with the bulk model with three monolayers embedded in a uniform dielectric background material, the plasmon dispersion for the coated structure suspended in air has plasmon branches which are slightly higher.

\section{Absorption}
\label{sec5}

When the layered heterostructure is subjected to an electromagnetic field carrying an electric polarization potential  $\Phi^{ext}({\bf  r})={\bf  E}_{ext}\cdot {\bf r}$, with frequency $\omega$ the vertical optical transitions from occupied to unoccupied states in the conduction band  will be stimulated. The absorption coefficient is determined from the number of transitions per incident flux of energy. Specifically, the absorption function for vertical transitions can be expressed in terms of the surface response function as \cite{Gumbs}

\begin{equation}
A(\omega)\propto  -\omega\ [1+ n _{ph}(\omega)]\Im  \int d{\bf q}\ 1/D({\bf  q},\omega+i\delta^+)\  ,
\label{absorption}
\end{equation}
where  $n _{ph}=1/[e^{\hbar\omega/k_BT}-1]$ is the photon distribution function.

\begin{figure}[h]
\centering
\begin{subfigure}{.4\textwidth}
\refstepcounter{subfigure}
\includegraphics[width=1.\columnwidth]{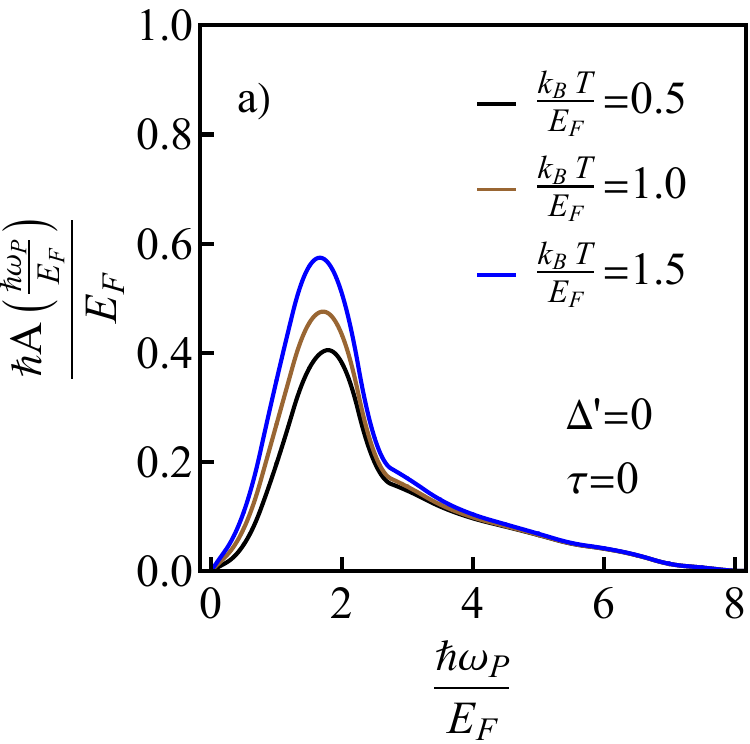}
\caption*{}
\label{abp1} 
\end{subfigure}%
\hspace{20pt}
\begin{subfigure}{.4\textwidth}
\refstepcounter{subfigure}
\includegraphics[width=1.\columnwidth]{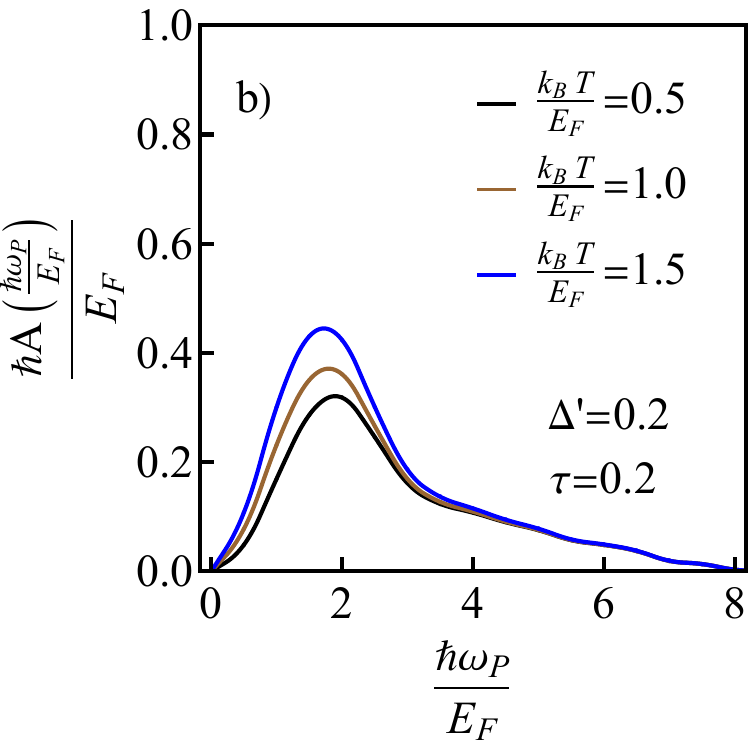}
\caption*{}
\label{abp2}
\end{subfigure}\captionsetup{singlelinecheck=off,justification=raggedright}
\caption{(Color online) Absorption function in unit of $E_F/\hbar$ with respect to the excitation frequency which is in unit of $E_F/\hbar$ for a monolayer of gapped TSDM on a substrate. Here, dielectric constant of the substrate is $\epsilon_1=5.0$. (a) Absorption coefficient when tilting $\tau=0$ and band gap parameter $\Delta'=0$. (b) Absorption when tilting $\tau=0.2$ and $\Delta'=0.2$.}
\label{absorption_mono}
\end{figure}

\medskip
\par
We plot the absorption function for a monolayer of gapped TSDM on a substrate for three different temperatures in Fig.~\ref{absorption_mono}. We observe that as the temperature is increased, the absorption is also increased. Also, compared to Fig.~\ref{abp1}, where tilting $\tau$ and the energy gap parameter $\Delta'$ are zero, when tilting and  $\Delta$ are applied with magnitude $0.2$, Fig.~\ref{abp2} shows that the absorption peaks are decreased.
\medskip
\medskip

\section{Conclusions}
\label{sec6}

In this paper, we derived a closed form analytic expression for the surface response function for a heterostructure consisting of up to three 2D layers. The layers are separated by dielectric media so that there is no electron tunneling between the layers or hybridization of the polarization functions. The resonances in the SRF correspond to the plasma excitations. The system could then be placed on a dielectric substrate and exposed to vacuum at the other end. From our general formula for three layers, we deduce  the dispersion equation for a heterostructure consisting of two 2D layers on the surface of a film of dielectric material and with a surface exposed to vacuum. This is in contrast to the case when two 2D layers are embedded in a bulk medium with uniform background dielectric constant. We compared these two cases (surface versus bulk) qualitatively and quantitatively and found that not only are the excitation frequencies affected by the homogeneous versus inhomogeneous background distribution of the dielectrics but the damping and separation between the plasmon branches are also affected in a nontrivial way.
\medskip
\par
The 2D structure which we chose to carry out our numerical calculations with is tilted anisotropic semi-Dirac material. This gave us  the opportunity to investigate effects due to the anisotropy and tilting of the energy bands on the plasmon dispersions all for one material. We presented long wavelength results for the plasmon dispersions for one and two layers as well as arbitrary wavenumber for trilayer, bilayer and monolayer TSDMs. These approximate long wavelength results agree well with the exact numerical values presented as density plots using the full dispersion function. The density plots do not only yield the numerical values for the plasmon excitation energies but also the strength of the damping in each branch. We find that the magnitude of the plasmon excitations is determined by tilting and the gap between the valence and conduction bands. For both the double and trilayer geometries, there are two and three  plasmon branches, respectively, in the absence of  interlayer hopping. Additionally,  the nature of the distributed background dielectric constant plays a crucial role on the visibility, closeness and even their anticrossing. We investigated these essential properties by comparing results obtained for a homogeneous (bulk) and inhomogeneous (surface) dielectric background and by varying the inter-planar separations. The plasmon branches differ in dampings from one branch to the next as well as along a chosen branch mode. For the double layer, if the charge density oscillations in both layers are in phase, the excitation energy is larger than the case when the oscillations are out of phase for either the bulk or when there is a surface. For the two layers, the damping of the higher optical branch is on the weakly damped side while that of the lower acoustic branch is on the highly damped side of the color codes. Furthermore, the plasmon dispersion of the optical modes obeys the bulk dispersion law $\omega_p\sim q_{x,y}^{1/2}$ whereas the acoustic branch satisfies $\omega_p\sim q_{x,y} $ in the long wavelength limit. For trilayer, the oscillations can all be in phase or two in-phase and the third out-of-phase which depends on the separation between the monolayers and also the dielectric background condition. The anisotropy of these modes is also revealed in two perpendicular directions  for the TSDMs with either a homogeneous or inhomogeneous dielectric background  for this layered structure. 
\medskip
\par
\medskip
\par
Based on our detailed analytical and numerical calculations for the plasma excitations in multi-layer 2D structures, it is abundantly clear that dielectric screening in conjunction with interlayer separation and the energy band structure can play a major role in applications to optoelectronics and photonics. These results identify dielectric screening as a means of tuning coupled plasmon modes without requiring interlayer electron tunneling or electronic hybridization, complementing earlier studies of Coulomb-coupled Dirac layers \cite{Profumo}. The dependence of plasmon dispersion and damping on wave-vector direction in tilted and gapped semi-Dirac systems further suggests opportunities for frequency- and direction-selective plasmonic responses, building on previous studies of anisotropic semi-Dirac plasmons \cite{Andrii}. A direct experimental application is to guide the interpretation of momentum-resolved EELS, which has resolved conductive and flexible, TSDMs can provide protection in key plasmon dispersion and spectral intensity in atomically thin $TaS_{2}$ monolayers and bilayers \cite{Do}. 
\medskip
\par
Our work may help to enable applications  to sensing, computing and energy storage through the manipulation of light on the nanoscale. This can be achieved by coupling light with plasmons. Now, with this nanoparticle consisting of several plasmon branches, we can have resonance of light at a given frequency with plasmons  at that frequency for several wavelengths. More broadly, the closed-form expressions provide an analytical reference for evaluating dielectric-environmental effects and benchmarking numerical calculations of multilayer plasmonic structures.

\medskip
\par

\begin{acknowledgements}
G.G. gratefully acknowledges funding from the U.S. National Aeronautics and Space Administration (NASA) via the NASA-Hunter College Center for Advanced Energy Storage for Space  under cooperative agreement 80NSSC24M0177. A.I was supported by the funding received from TradB-56-75, PSC-CUNY Award 68386-00 56. We also acknowledge support from the Science and Technology Facilities Council (STFC), UK (Reference No. ST/Y005147/1). The views expressed are those of the author and do not necessarily reflect the official policy or position of the Department of the Air Force, the Department of Defense, or the U.S. government.
\end{acknowledgements}

{}

\end{document}